\documentclass{aa}
\usepackage{graphicx}

\begin{document}
\title{A dynamical model for FR\,II type radio sources with terminated jet activity}
   
\author{El\.zbieta Kuligowska
   }

\offprints{E. Kuligowska \email elzbieta@oa.uj.edu.pl}

\institute{Astronomical Observatory of the Jagiellonian University,
    Orla 171 Cracow, Poland }

\date{Received 15 June 2016 / Accepted 30 June 2016}

 \abstract {The extension of the KDA analytical model of FR\,II$-$type source evolution 
originally assuming a continuum injection process in the jet$-$IGM (intergalactic medium) interaction 
towards a case of the jet’s termination is presented and briefly discussed.}
{The dynamical evolution of FR\,II$-$type sources predicted
with this extended model, hereafter referred to as KDA\,EXT, and its application to the chosen radio sources.}
{Following the classical approach based on the source's continuous injection and self-similarity, 
I propose the effective formulae describing the length and luminosity evolution of the lobes during an
absence of the jet flow, and present
the resulting diagrams for the characteristics mentioned.}
{Using an algorithm based on the numerical integration of a modified formula for jet power,
the KDA\,EXT model is fitted to three radio galaxies. Their predicted
spectra are then compared to the observed spectra, proving that these fits are better than the best spectral 
fit provided by the original KDA model of the FR\,II$-$type sources dynamical evolution.}
{}

\maketitle 

\section{Introduction}
The previously elaborated and published analytical models for the dynamics 
and radio emission properties of FR\,II$-$type radio sources (Fanaroff \& Riley 1974)
are all based on the "standard model" of Scheuer (1974) and Blandford \& Rees
(1974); the model explains these sources as an interaction of twin jets, composed of highly
relativistic particles emerging from the active galactic nucleus (AGN), with 
the external gaseous environment called intergalactic medium (IGM) surrounding a given galaxy or quasar.
The jet formation results from the matter's accretion processes on a supermassive 
black hole with magnetic field and angular momentum. Interaction
of the jets with the IGM produces a supersonic shock enabling the acceleration of the external 
medium particles and the jet material after their transition through
the shock which, in turn, inflates the radio lobes or cocoon and is observed as
diffuse emission with a typical power-law spectrum.

Non-thermal continuum radio emission of the lobes is due to the 
synchrotron process and to the inverse-Compton scattering of ambient photons
of the cosmic microwave background (CMB). The synchrotron radiation arises from
ultra-relativistic particles interacting with the magnetic field. According to 
classical electrodynamics, the life$-$time of relativistic electrons emitting medium
and short radio waves has to be relatively short. The most energetic electrons,
preferentially emitting at high radio frequencies and due to the synchrotron losses,
lose their energy at the fastest rate. This implies that, in the absence of a
constant injection of fresh particles into the lobes, the synchrotron losses must
cause a violent steepening of their radio spectra and a rapid dimming of the
source's structure.

However, observed radio lobes are stable, so such long-lived structures have
to be continuously powered by new particles from the AGN. This scenario
forms the basis of all analytical models of powerful radio galaxies and radio-loud
quasars (e.g. Begelman \& Cioffi 1989, Falle 1991, Kaiser et al. 1997,
Blundell et al. 1999, Manolakou \& Kirk 2002). All of these models, which differ in
their description of how the relativistic particles pass from the jets into the lobes and 
in their treatment of energy losses and particle transport, assume a constant
injection of new particles to the lobes during the time at which the source is observed.
This is, clearly, a substantial simplification of these models.

Nevertheless, precise observations show several radio structures (of FR\,II$-$type)
with so-called double-double morphology strongly suggesting a recurrent jet
activity in the AGN (cf. Saikia \& Jamrozy 2010) and also show radio spectra of the lobes
with a high-frequency slope significantly exceeding the limiting value of
$\alpha_{\rm inj}+0.5$ expected in the continuum-injection (C.I.) process.
Therefore, in this paper I modify the model of Kaiser, Dennett-Thorpe \&
Alexander (1997; hereafter KDA model) by introducing one more free parameter
in that model, i.e. a time of the jet's termination, $t_{\rm br}$, and consider
changes in the adiabatic expansion of the cocoon (lobes) and in its radio
emission at $t_{\rm br}$.

This approach was already extensively discussed
by Kaiser et al. (2000) in their investigation of the evolution of the outer and
inner structures of five selected double-double radio galaxies. However, their
considerations were qualitative rather than quantitative. The quantitative
results given in the above paper were reached under a number of arbitrary
assumptions for the values of basic and very sensitive free parameters of the
model, fixing the value of the density of the radio core, fixing the value of the 
exponent of the initial energy distribution of the relativistic particles in the jet
flow, taking the unique value of the aspect ratio of the cocoon, etc. Our aim is
to provide the model that is flexible enough, thus enabling the determination of these 
parameters by the fit of the extended model to the observational data of a 
possibly large representative sample of FR\,II$-$type radio sources selected
within a wide range of linear size, radio power, and redshift.   

Therefore, in this paper I propose the effective formula for describing the dynamics
and the luminosity evolution of the lobes of FR\,II$-$type sources during an absence
of the jet inflow. The brief summary of the original KDA model and its proposed
extension is given in Section\,2. The $P_{\rm \nu}-D$ diagrams, radio spectra, and other characteristics of
dynamical evolution of FR\,II$-$type radio sources predicted with this extended model 
are analysed in Section\,3. The model fits to the observational
data for the three example radio galaxies, providing the expected values 
of their basic physical parameters (the jet power, internal density and pressure,
magnetic field strength, the source’s age, and duration of the jet flow) are
presented in Section\,4 along with the radio spectra of these sources resulting from the new model. The goodness 
of the fit of these spectra to the observational 
data and the corresponding fits arising from the original KDA model are included. 
The discussion of the results is presented in Section\,5.

\section{Applied model and its extension}
\subsection{Brief summary of the original KDA model}

The model applied is based on a detailed dynamical description by Kaiser
\& Alexander (1997) combined with the radiative processes analysed by
Kaiser et al. (1997) and extensively summarized in Barai \& Wiita (2006).
It assumes a power-law radial density distribution (the simplified King 1972 profile) 
of unperturbed ambient gas surrounding the radio source as
$\rho_{\rm a}(r)=\rho_{0}(r/a_{0})^{-\beta}$, where $\rho_{\rm a}$ is the ambient
density at distance $r$ from the centre of the host galaxy, $\rho_{0}$ is the central
density of the radio core with radius $a_{0}$, and $\beta$ is exponent of the
density profile. From the energy conservation conditions arises the expression for the total
length of the jet at a given age $t$ as

\begin{equation}
r_{\rm j}(t)=c_{1}\left(\frac{Q_{\rm j}}{\rho_{0}a_{0}^{\beta}}\right)
^{1/(5-\beta)}t^{3/(5-\beta)},
\end{equation}
where $Q_{\rm j}$ is the jet’s power and $r_{\rm j}$ is identified with one-half
of the source’s linear size $D$, $r_{\rm j}=D/2$. If two of the model parameters, 
$Q_{\rm j}$ and $\rho_{0}a_{0}^{\beta}$, are specified, the model
predicts the time evolution of the source (its size $D$ and volume of the
cocoon, $V$) only.

The independent relation between the above parameters is available from 
a consideration of expected radio emission of the source (its lobes or cocoon)
under influence of different energy losses. The energy loss of an electron with
the Lorentz factor $\gamma$ is
\begin{equation}
\frac{d\gamma}{dt}=-\frac{a_{1}}{3}\frac{\gamma}{t}-\frac{4}{3}
\frac{\sigma_{\rm T}}{m_{\rm e}c}\gamma^{2}(u_{\rm B}+{\rm iC}),
\end{equation}
where the first term on the right-hand side refers to the adiabatic expansion loss in the
source’s volume evolving as $V(t)\propto t^{a_{1}}$ and the second term to the
combined energy loss due to the synchrotron radiation and inverse-Compton
scattering of the CMB radiation; $u_{\rm B}$
and ${iC}$ are the energy density of the magnetic field and CMB photons,
$a_{1}=(4+\beta)/{\Gamma_{\rm c}(5-\beta)}$. Moreover, this model assumes
a constant rate at which relativistic particles in the jets are transported from the
AGN to the hot spot area interpreted as the region of the particles’ reacceleration.
The initial energy distribution of injected particles is taken as 
$n(\gamma_{\rm i})=n_{0}\gamma_{\rm i}^{-p}$, where $p$ is constant.
Minimum energy arguments give the ratio $r$ of the magnetic field’s energy
density to the sum of the energy densities of the relativistic, $u_{\rm e}$,
and thermal, $u_{\rm T}$, particles:
$r=u_{\rm B}/(u_{\rm e}+u_{\rm T})=u_{\rm B}/[u_{\rm e}(1+k')]=(1+p)/4$.

Finally, the radio emission of the source (cocoon) at a given frequency is
calculated by splitting it into infinitesimal evolving volume elements, i.e.
undergoing the adiabatic and radiative losses (cf. Equation\,2). The sum of the
contributions from all of these elements gives the total emission (at a
frequency $\nu$), $P_{\nu}(t)$, as a complicated integral over injection
time $t_{\rm i}$

\[
P_{\nu}(t) = 
\int\limits^t_{t_{\rm min}} \, dt_{\rm i} \frac{\sigma_{T}c\;r}{6{\pi}\nu(r+1)} 
Q_{\rm j}n_{0}(P_{\rm hc})^{(1-\Gamma_{\rm c})/\Gamma_{\rm c}}
\]
\vspace{-4mm}
\begin{equation}
 \times\frac{\gamma^{3-p}t_{\rm i}^{a_{1}/{3(p-2)}}} 
{[{t^{-a_{1}/3}}-a_{2}(t,t_{\rm i})\gamma)]^{2-p}} 
 \left(\frac{t}{t_{\rm i}} \right)^{-a_{1}(1/3+\Gamma_{\rm B})},
\end{equation}
\vspace{2mm}
\noindent  
where $(P_{\rm hc})$, the ratio of the jet head pressure $(p_{\rm h})$ and 
the uniform cocoon pressure $(p_{\rm c})$, is a function of $(R_{\rm T})$ - 
the axial ratio of the cocoon described by the empirical formula adopted 
from Kaiser (2000); $\Gamma_{\rm c}$ and  $\Gamma_{\rm B}$ are the adiabatic indices 
in the equation of state of the cocoon material and the magnetic field, respectively. 

The integration of Equation\,3 is performable using a numerical calculation only. 

\subsection{Extension of the KDA model}

\subsubsection{Adiabatic expansion of the cocoon after nuclear activity ceases}

The information about termination of the energy supply propagates
from the AGN to the radio lobes with the speed of sound. There are
arguments for a relatively low internal sound speed in the lobes, for
example if there is a significant mixing of the lobe material with the
surrounding gaseous environment (e.g. Kaiser et al. 2000). It implies that
after switching off the jets, the adiabatic evolution of the lobes of old
and large sources may be the same as it was before cease of the
nuclear activity for a long time. However, for small and overpressured
(with respect to the external medium) lobes, the internal sound speed
can be high and their adiabatic evolution can slow down in a relatively
short time. 

After switching off the supply of new particles to the cocoon, it may still
be overpressured with respect to the external gaseous environment and
therefore be continuing its expansion behind a bow shock. Such adiabatic
evolution of a spherical source (early evolution of a supernova remnant)
during the so-called coasting phase was analysed by Kaiser \& Cotter
(2002). Although the cocoons of radio sources and the
related bow shocks are not spherical, they assume a spherical
bow shock with the radius $R_{\rm S}$ growing with time as
$R_{\rm S}\propto t^{2/(5-\beta)}$. For simplicity they studied the
scenario where a radio source reaches pressure equilibrium with
the external medium immediately after the jet switches off and assumed
an instantaneous transition between the active and coasting phases.
With these assumptions they solved the equation of state for the spherical
approximation of the source resulting in steady-state similarity
solutions from which we can identify their source’s radius $R_{\rm c}$
with the lobe length $D/2$ in the KDA model. After some transformations, 
the effective formula for the cocoon (the source) length, describing its
adiabatic evolution before and after termination of the jet's activity, is
given by

\begin{equation}
D(t, t_{\rm br})= \left \{{
c_{1}\left(\frac{Q_{\rm j}}{\rho_{0}a^{\beta}_{0}}\right)
^{1/{(5-\beta)}}t^{{3/{(5-\beta)}}} \hspace{20pt} for\,t < t_{\rm br}  \atop { 
D(t_{\rm br})\left(\frac{t}{t_{\rm br}}\right)
^{\frac{2(\Gamma_{\rm c}+1)}{\Gamma_{\rm c}(7+3\Gamma_{\rm c}-2\beta)}} \hspace{17pt} 
for\,t \geq t_{\rm br}}}\right.
\end{equation}
where 
$D(t_{\rm br})$ is the cocoon length at the time of switching off the energy
supply. Hereafter we follow the Kaiser \& Cotter approach in the limiting
case when the internal sound speed is fast (their model B).

\subsubsection{Analytical formula for the integration of radio power}

The energy loss process due to the synchrotron and inverse-Compton
scattering of the CMB photons (second term on the right-hand side in Equation\,2) 
is characterized by the energy break in the energy spectrum comprising
particles with different energies. Expressing the energy densities in
Equation.\,2, $u_{\rm B}$ and $u_{\rm iC}$, by the corresponding magnetic
field strengths $B$ and $B_{\rm iC}$ and the constant
coefficient by the relevant constant defined by Pacholczyk (1970), one
has
\begin{equation}
\frac{d\gamma}{dt} = -C_2\{(B\sin \theta)^2+B_{\rm iC}^2\} \gamma^{2},
\end{equation}
where $\theta$ is the pitch angle of the relativistic particles.
Integrating Equation\,5 gives the time evolution of the particles’ energy
$\gamma(t)$ with the energy break $\gamma_{\rm br}^{-1}=
C_{2}\{(B\sin\theta)^{2}+B_{\rm iC}^{2}\}^{2}t$ 
fulfilling the condition $\gamma(t)/\gamma_{\rm br}<1$. The appearance of
$\gamma_{\rm br}$ causes a break frequency in the observed radio
spectrum. If the source is supplied by a constant flow of particles
(i.e. the continuous injection process; hereafter CI), the highest break
frequency (for $\theta=90\degr$, i.e. for the particles dominating the
emission) is
\begin{equation}
\nu_{\rm br,CI}=C_{1}B\gamma_{\rm br}^{2}=
\frac{C_{1}}{C_{2}^{2}}\frac{B}{(B^{2}+B_{\rm iC}^{2})^{2}t^{2}}
\end{equation}
\noindent
which we assume is valid for $t\leq t{\rm br}$, and 
$C_{1}$ and $C_{2}$ are the physical constants (cf. Pacholczyk
1970). 
On the other hand, if the pitch angle of the particle distribution is
highly isotropic, i.e. all the particles can have any possible pitch
angles (the Jaffe-Perola process; hereafter JP), the break frequency
valid for $t > t{\rm br}$ is  
\begin{equation}
\nu_{\rm br,JP}=\frac {C_{1}}{C_{2}^2} \frac{B}{\{(2/3B^{2}+B_{\rm iC}^2)(t-t_{\rm br})\}^2}
\end{equation}
where $C_{1}/C_{2}^{2}=2.51422\times 10^{12}$ and $B_{\rm iC}=
0.318(1+z)^{2}$ nT.

\vspace{2mm} 
According to the superposition principle the analytical formula for the
total radio power of a source (its cocoon) at a given frequency can be
rewritten as the sum of two integrals. In the extended model presented
here, the first integral gives the source power calculated until the time
of the jet termination, $t_{\rm br}$, while the second adds the radio
power emitted from $t_{\rm br}$ until the actual age of the source, $t$.
Therefore, the total power of a source can be written as

\begin{equation}
P_{\rm \nu}(t)= \left \{{
P_{\rm\nu}(t_{\rm min},t_{\rm br}) + P_{\rm\nu}(t_{\rm br},t)\hspace{25pt} for \hspace{5pt}t_{\rm br} > t_{\rm min} } \atop { 
P_{\rm\nu}(t_{\rm min},t)\hspace{84pt} for \hspace{5pt}t_{\rm br}\leq t_{\rm min}}\right.
\end{equation}
In the above equation the first term corresponds to the integral given
by Equation\,3 in Section\,2.1, where the upper limit of integration is changed
from $t_{\rm min}$ to $t_{\rm br}$. The second term is given by

\begin{eqnarray}
P_{\rm\nu}(t_{\rm br},t) = \frac{\sigma_{T}c}{6\pi\nu}\frac{r}{r+1}Q_{\rm j}P_{\rm hc}^\frac{(1-\Gamma_{c})}
{\Gamma_{c}}\int\limits^{t}_{t_{*}}\,G(t_{\rm i})H(t_{\rm i})dt_{i}, 
\end{eqnarray}

\noindent
where $t_{*}$ = $t_{br}$ if $t_{\rm br} > t_{\rm min}$ and $t_{*}$ = $t_{\rm min}$ if $t_{\rm br} \leq t_{\rm min}$,

\[
H(t_{\rm i}) = n_{0}(t_{\rm i})\frac{\gamma^{3-p}t_{\rm i}^{a_{1}/3(p-2)}}{\{t^{-a_{1}/3}-a_{2}(t,t_{\rm i})\gamma\}^{2-p}} \left
( \frac{t}{t_{\rm i}}\right)^{-a_{1}(1/3+\Gamma_{\rm B})},
\]

%\[
%G(t_{i}) = \frac{\int\limits^{1}_{0}\,F(\frac{\nu/\nu_{\rm br,JP}}{x^2}) x^{-p}(1-x)^{p-2}dx}
%    {\int\limits^{1}_{0}\,F(\frac{\nu/\nu_{\rm br,CI}}{x^2}) x^{-(p+1)} \{1-(1-x)^{p-2}\}dx + \int\limits^{\infty}_{1}\,F(\frac{\nu/\nu_{\rm br,CI}}{x^2})x^{-(p+1)}dx}, \]
\[
G(t_{\rm i})=
\]

\vspace{-12mm}

\[
\frac{\int\limits^{1}_{0}\,F_{\rm JP}(x) x^{-p}(1-x)^{p-2}dx}
    {\int\limits^{1}_{0}\,F_{\rm CI}(x)x^{-(p+1)} 
\{1-(1-x)^{p-2}\}dx+\int\limits^{\infty}_{1}\,F_{\rm JP}(x)x^{-(p+1)}dx}, \]

\noindent
and the functions $F_{\rm JP}=({\nu/\nu_{\rm br,JP}/x^2)}$ and $F_{\rm CI}=(\nu/\nu_{\rm br,CI}/x^2)$ are 
the ’kernel’ synchrotron spectrum of a single ultra-relativistic particle. 
It should be noted that transient values of
$\nu_{\rm br,JP}$ and $\nu_{\rm br,CI}$ are functions of $t_{\rm i}$,
i.e. the time of initial acceleration of the radiating particles. The sum
of the two integrals in the denominator of the function $G(t_{\rm i})$
results from a different form of these integrals for $\nu<\nu_{\rm br}$
and for $\nu\geq \nu_{\rm br}$. It is calculated at every step of the
integration and all of the infinitesimal values of integrated radio
power are multiplied by this function.

\section{Predictions of the extended model}
~ 
In order to study the influence of the jet termination 
on the $P_{\nu}-D$ diagrams and observed radio spectra, a fiducial source with a set of 
basic model parameters: $\alpha_{\rm inj}$=(p-1)/2=0.51, $\beta$=1.5, $z$=0.1, $R_{t}$=3, $k'$=0, adiabatic indices in the 
equation of state: $\Gamma_{\rm c}$=$\Gamma_{\rm B}$=5/3, $a_{\rm 0}$=10 kpc, $Q_{\rm j}$=$10^{38}$ 
W and $\rho_{0}$=$10^{-22}$ \rm $\rm {kg\,m}^{-3}$ is used.

\subsection{Evolutionary tracks}
~
Kaiser, Denett-Thorpe \& Alexander (1997) presented the $P_{\rm \nu}-D$ diagrams (Shklovskii 1963) 
for the original KDA model applied to fiducial FR\,II$-$type radio sources with parameters
adopted in agreement with the present knowledge. These diagrams, which constitute a powerful tool for 
investigation of time evolution of radio sources, show the sources decreasing radio luminosity determined at a 
given frequency, $P_{\rm \nu}$, as a function of its linear size $D$ enlarging with time.  
In this section, the analogous evolutionary tracks of the fiducial FR\,II$-$type sources, 
constructed with the new KDA\,EXT model and with the use of some revised physical parameters of the source 
(cf. the list given above) are presented.
 
Figure\,1 shows the $P_{\rm 178}$ - $D$ diagrams for the fiducial source with three different values of $\alpha_{\rm inj}$=
0.51, 0.75, and 1.0 for $t_{\rm br}$=10 Myr. The dashed and solid lines show its lobes radio power and size evolution 
with time in two limiting scenarios: a very slow and a fast internal sound speed in the 
cocoon, respectively (see Section\,2.2.1). Within the first scenario, the lobe's length is initially still 
evolving as in the continuum-injection model (the upper expression of Equation\,4), and after some time it begins to evolve 
as in the KDA\,EXT model.
In the second scenario, lobes are continuously expanding according to the KDA\,EXT model and formula described by the bottom 
expression of Equation\,4. Finally, dotted lines correspond to the pure KDA evolution. 

These diagrams demonstrate the dependence of various source parameters on its dynamical
evolution in the two extreme cases of the internal sound speed (KDA and KDA\,EXT). 
The clearly visible breaks may result from the cocoon length evolution (here adopted from Kaiser \& Cotter's limiting
case B, in which the internal sound speed is fast - cf. Section\,2.2.1) is not completely applicable in the case of
terminated jet activity. In particular, the information about stopping the jet is not transmitted instantaneously, so the
lobes should not begin to expand significantly more slowly immediately after the jet switch-off. It is more probable that its evolution 
changes smoothly from one regime (CI) to another (KDA\,EXT). 

Since we are not able to provide the right formula for this gradual change, we can only approximate this 
situation by using the following procedure based on the
assumption that the rate of the cocoon growth only changes slowly: the separated model is constructed with the radio luminosity evolving as 
in the KDA\,EXT model, but with the lobe length still growing as in the CI regime (KDA) shortly after the jet termination. Then, after the assumed time, the lobes length 
evolution transforms into the internal sound speed scenario of KDA\,EXT, but with more recent stopping of the jet activity, which implies using the 
reduced value of $t_{\rm br}$ (here equal to the 0.7 of the initial $t_{\rm br}$ in the KDA\,EXT). This procedure simulates - very roughly - 
the slow change described above and is intended to demonstrate that the 'broken' KDA\,EXT plots can be 
easily smoothed only if the proper formula for the gradual lobe growth can be applied.

It is also worth noting that the abnormal breaks on $P_{\rm 178}$ - $D$ diagrams may be 
the result of numerical effects occurring in the calculation of the lobes radio luminosities. In particular, they 
may occur because the radio power in the KDA\,EXT model constitute the sum of three different and independently calculated integrals with the distinct 
integration limits (cf. Equation.\,8). The characteristic 'shoulder' visible in all diagrams corresponds to time $t_{\rm i}$
at which the integration changes from the case of $t_{\rm br}$ $>$ $t_{\rm min}$ to 
$t_{\rm br}\leq t_{\rm min}$. This implies that though we expect the plots to be continuous at this point, 
some breaks appear as the numerical result of changing the limits of integration and (most likely) more rapid decrease in the first of 
the integrals in the upper expression of Equation\,8, which cannot keep up with the slower increase in the second integral in this 
formula and maybe also with the independently evolving lobe length. It is also interesting that these breaks are sharper for the high values of $\alpha_{\rm inj}$ 
(cf. Figure\,1)

Figure\,2 shows the $P_{\rm 178}$ - $D$ diagrams for three different values of $\beta$: 0, 1.5, and 1.9 and for 
$\alpha_{\rm inj}$=0.51 and for $t_{\rm br}$ = 10, 20, and 100 Myr, respectively.  
The solid, dashed, and dotted lines correspond to the same models as in Figure\,1.
Finally, Figure\,3 present analogous $P_{\rm 178}$ - $D$ diagrams for the fiducial source for three different 
values of $\rho_{0}$: $10^{-21}$, $10^{-22}$, and 
$10^{-23}$ \rm kg m$^{-3}$, with $\alpha_{\rm inj}$=0.51 and for $t_{\rm br}$=10 Myr. 
Regardless of varying parameters of the fiducial source, evolutionary tracks for the KDA\,EXT model 
behave similarly to the ones discussed above, and also here the characteristic plot irregularities can be noticed. It can be then 
assumed that they are also most probably due to the numerical effects or incorrect formula for calculating 
the lobe length.

\begin{figure}[h]
\includegraphics[width=11cm, height=13cm]{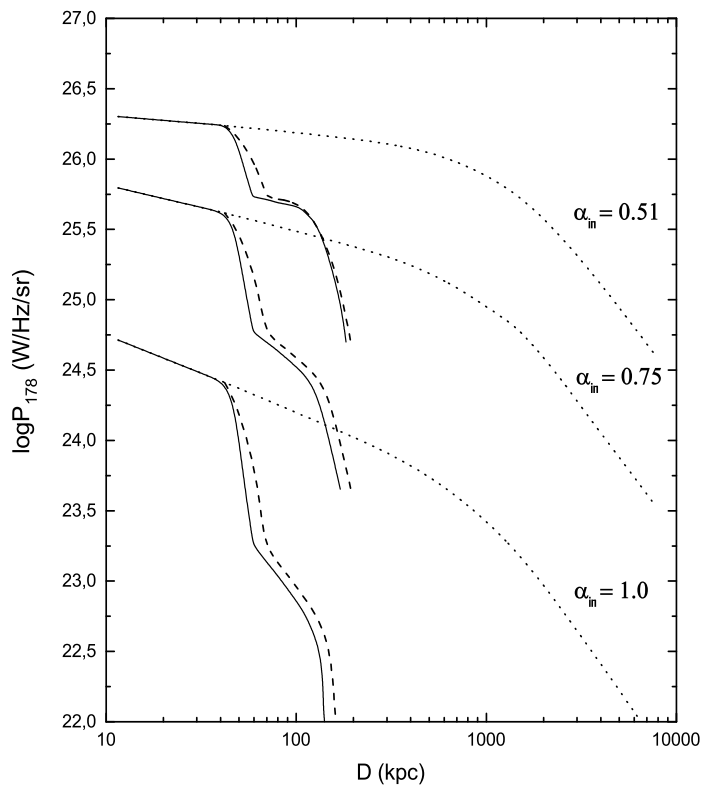}
\caption{Comparison of the $P_{178}-D$ diagrams predicted with the KDA and 
KDA\,EXT model for the fiducial source with three different values of $\alpha_{inj}$ and $t_{\rm br}$ = 10 Myr. 
The solid and dashed lines show predictions of the KDA\,EXT model for the two limiting cases of the internal 
sound speed (see the text). 
The dotted lines show predictions of the pure KDA model. All the KDA\, EXT diagrams are cut off at the points where their radio powers 
rapidly diminish and linear sizes tend to a constant value.}
\label{1}
\end{figure}

\clearpage

\begin{figure}
\includegraphics[width=11.05cm, height=13.05cm]{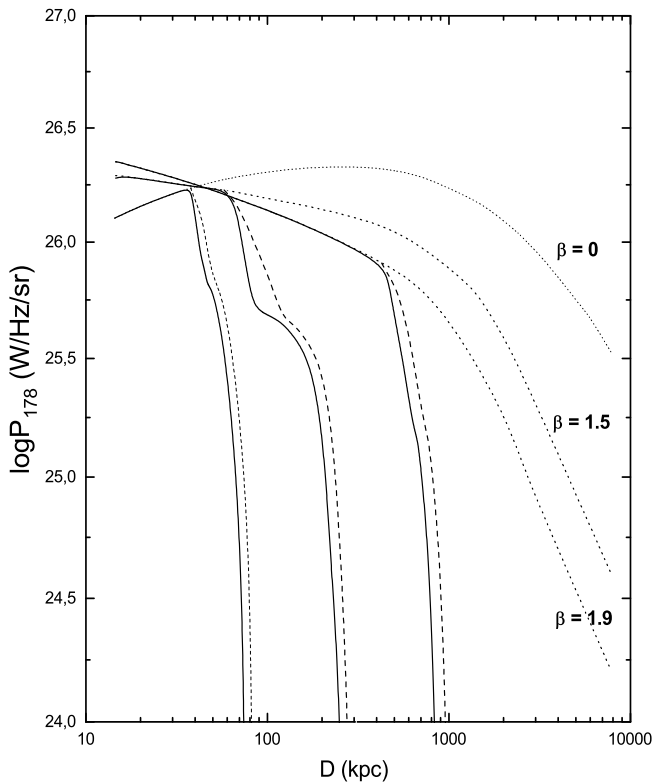}
\caption{Comparison of the $P_{178}-D$ diagrams predicted by the KDA and KDA\,EXT model with 
different values of $\beta$ (0, 1.5, 1.9) and $t_{\rm br}$ = 10, 20, and 100 Myr, respectively. 
Calculations are performed for the same set of source parameters as 
in Fig.\,1, but only for $\alpha_{inj}$=0.51. Solid lines show predictions of the KDA\,EXT, 
dotted lines show predictions of the KDA model, while the dashed lines trace the corrected (smoothened) KDA\,EXT model with 
the approximated gradual change of the lobe length evolution (see the text).}
\label{2}
\end{figure}

\begin{figure}
\centering
\includegraphics[width=11cm, height=13cm]{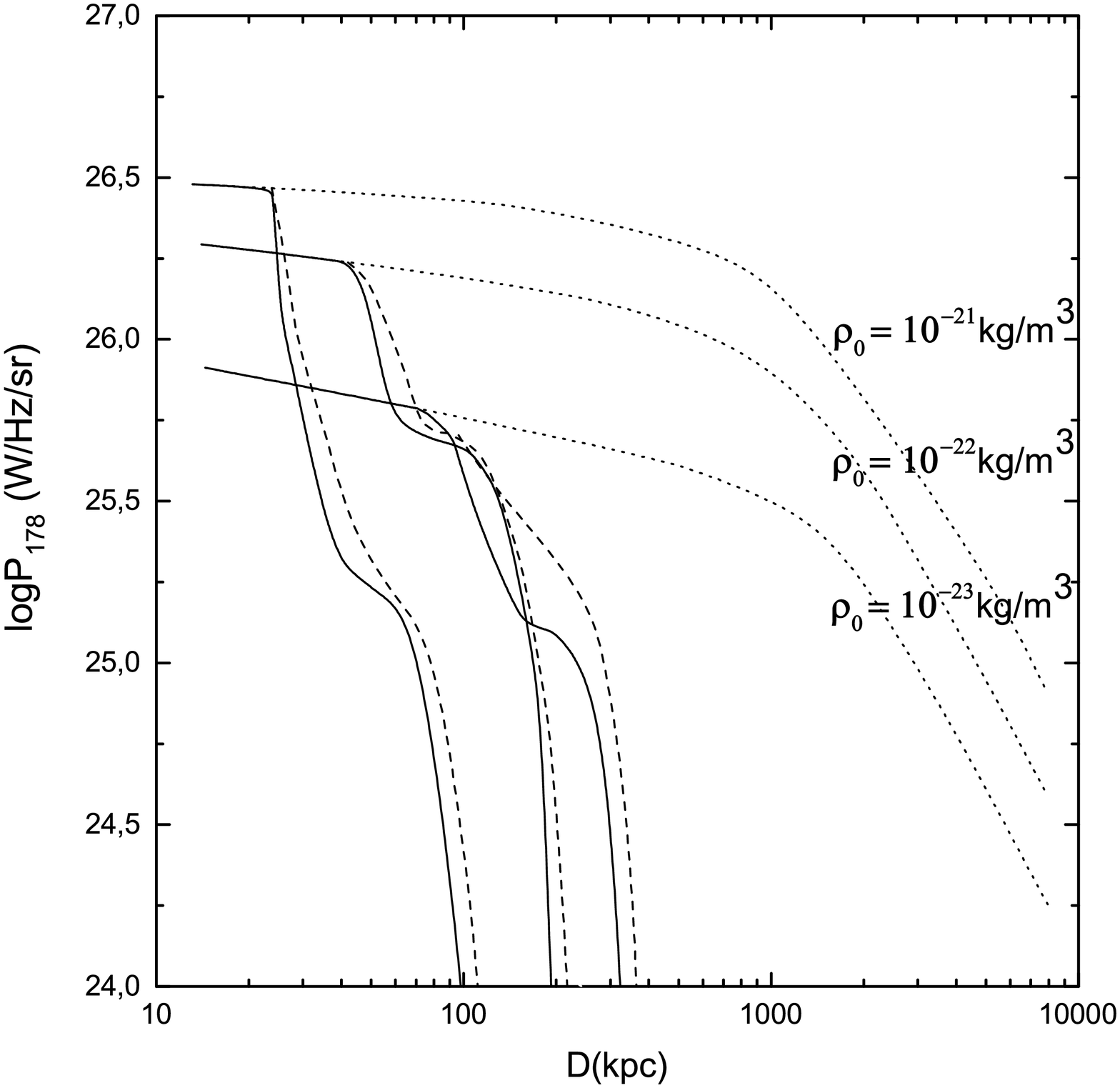}
\caption{Comparison of the $P_{178}-D$ diagrams predicted by the KDA and KDA\,EXT model with 
different values of $\rho_{0}$ and $t_{\rm br}$ = 10 Myr. 
Calculations are performed for the same set of source's parameters as 
in Fig.\,1, but only for $\alpha_{inj}$=0.51. Solid lines show predictions of the KDA\,EXT, 
dotted lines show predictions of the KDA model, and the dashed lines trace the same predictions as the analogous lines in the Figure\,2.}
\label{3}
\end{figure}

\subsection{Time evolution of the lobe's expansion velocity, internal pressure and the total energy emitted}
~
The relations between the source's age and the expansion velocity of its lobes, the 
internal pressure and total energy emitted are also analysed. The instantaneous bow shock velocity $v_{h}$ is

\begin{equation}
v_{\rm h}(t) = \frac{\partial}{\partial t}D(t)
\end{equation}

\noindent
and is reduced to the differentiation of Equation\,4 which gives:

\begin{equation}
v_{\rm h}(t)= \left \{{
c_{1}{\frac{3}{5-\beta}}\left(\frac{Q_{j}}{\rho_{0}a^{\beta}_{0}}\right)
^{\frac{1}{5-\beta}}t^{\frac{\beta-2}{{5-\beta}}},\hspace{16pt}   t < t_{\rm br}  \atop 
{c_{4}D(t_{\rm br})t_{\rm br}^{-c_{4}}t^{c_{4}-1},\hspace{36pt}}  t \geq t_{\rm br}}\right.,
\end{equation}

\noindent
where
$c_{4}={\frac{2(\Gamma_{\rm c}+1)}{\Gamma_{\rm c}(7+3\Gamma_{\rm c}-2\beta)}}$.

\vspace{3mm}

Figures\,4 - 5 present the evolution of the lobe's head velocity 
given as the ratio of the speed of light, $c$, in function of the source's age for the KDA 
and the KDA\,EXT model. It is worth noting that the rate of its decrease 
depends not only on $t_{\rm br}$ and the $\beta$ exponent, but also on different 
values of $\Gamma_{c}$ (for relativistic and cold cocoon material). The diagrams
indicate that the process of termination of the nuclear activity strongly affects the rate
of the lobe propagation as well. The difference of this rate increases with the increase in the 
value of the $\beta$ exponent describing the density profile of the external medium.

In the case of a young source, the lobe expansion velocity predicted with the KDA
model strongly depends on the density profile of that medium (cf. Figure\,5).
For the radio sources with terminated activity (the case of KDA\,EXT model) this dependence rapidly
declines with their growing age because the $\beta$ parameter has a relatively small
contribution in the formula describing the lobe's head velocity. On the contrary, this
velocity begins to depend more strongly on the value of $\Gamma_{c}$ with the source's age. It can 
also be seen that for the assumed value of $\beta$=0, the lobe velocities

\clearpage

\begin{figure}[h]
\includegraphics[width=10cm, height=10cm]{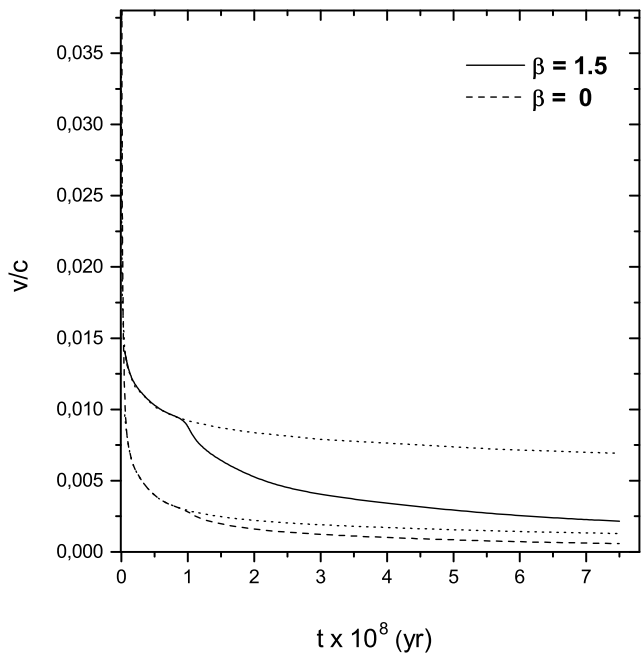}
\caption{Velocity of the expanding lobes vs. time. Solid and dashed curves correspond to the 
KDA\,EXT and dotted lines follow the KDA model of continuous
activity. Parameters are the same as in Fig.\,1, but with $t_{\rm br}$=100
Myr and $\Gamma_{c}$=5/3.}
\label{4}
\end{figure}

\begin{figure}
\includegraphics[width=10cm, height=10cm]{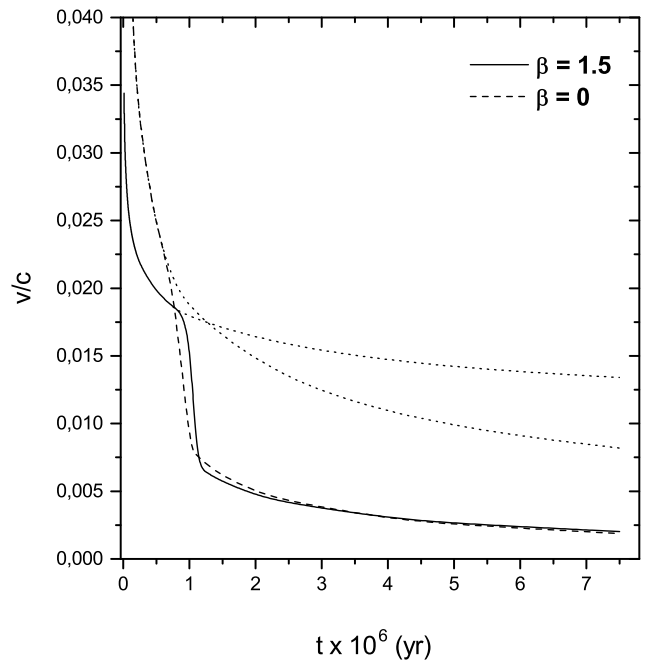}
\caption{Velocity of the expanding lobes vs. time. As in Fig.\,4, but
for $t_{\rm br}$=1 Myr.}
\label{5}
\end{figure}

\noindent
predicted from the models with different values of $t_{\rm br}$ are very similar. The difference between $v_{\rm}$
relations is much more explicit in the case of these two models calculated for
$\beta$=1.5.

Following the discussion given by Kaiser \& Cotter for their model B, I also assume
that the time evolution of the lobe pressure $p_{\rm c}$ after switching off the
source's nuclear activity, changes its former decrease according to the formula:

\begin{equation}
p_{\rm c}(t > t_{\rm br}) = p_{\rm c}(t_{\rm br})\left(\frac{t}{t_{\rm br}}\right)^{-3\Gamma_{\rm c}c_{4}}.
\end{equation}

Figure\,6 presents the behaviour of internal pressure of the expanding lobes due to its growing age. 
Similarly to the lobe's head velocity, the
evolutionary tracks are calculated for both the KDA and KDA\,EXT models for the two different 
values of the $\beta$ exponent. In both models the slope of the $p_{c}(t)$ function is steeper
for $\beta$=1.5 than for $\beta$=0, according to Equation\,19. Regardless of that
factor, the time evolution of the lobe's internal pressure is faster and much more rapid
in the KDA\,EXT model, which is in good agreement with our physical intuition - in the
case of the termination of the jet, the decrease in the lobe's head velocity is expected, as is 
the gradual decline in that pressure. It is also noteworthy
that in the case of the KDA\,EXT, there is an exponential decay of the
internal pressure that, starting at the particular age, is independent of the $\beta$
value. Finally, the evolution of the lobe's internal pressure depends very weakly on
the value of $\Gamma_{c}$.

Figures\,7 and 8 show the conversion efficiency of the energy delivered by the jet until its termination 
into the observed radiation as a function of $t/t_{\rm br}$. We can expect its decrease 

\begin{figure}
\includegraphics[width=10cm]{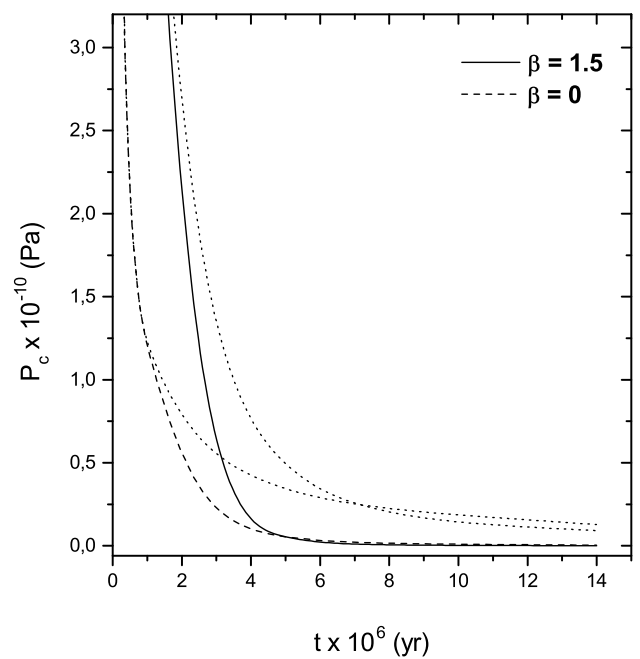}
\caption{Internal pressure of the expanding lobes vs. time. Solid and dashed curves correspond 
to the KDA\,EXT and dotted lines follow the KDA model of
continuous activity. The parameters are the same as in Fig.\,1, with
$t_{\rm br}$=1 Myr and $\Gamma_{c}$=5/3.}
\label{6}
\end{figure}

\clearpage
\noindent 
owing to the fiducial source's growing age.
Similarly to the lobe's head velocity, the evolutionary tracks are calculated for both KDA and KDA\,EXT models 
for three different values of the $\beta$ exponent 
and three values of $R_{\rm T}$. Both diagrams are presented: $\Gamma_{c}$=4/5 and $\Gamma_{c}$=5/3. The slopes 
of the efficiency function are steeper for the low values of $\beta$ and 
$R_{\rm T}$, and its evolution is much faster for sources with terminated activity while its value is constant during the phase 
before the jet termination. The value and the further evolution of the lobe's efficiency depends strongly on the value of $\Gamma_{c}$.

\begin{figure}
\includegraphics[width=10cm, height=10cm]{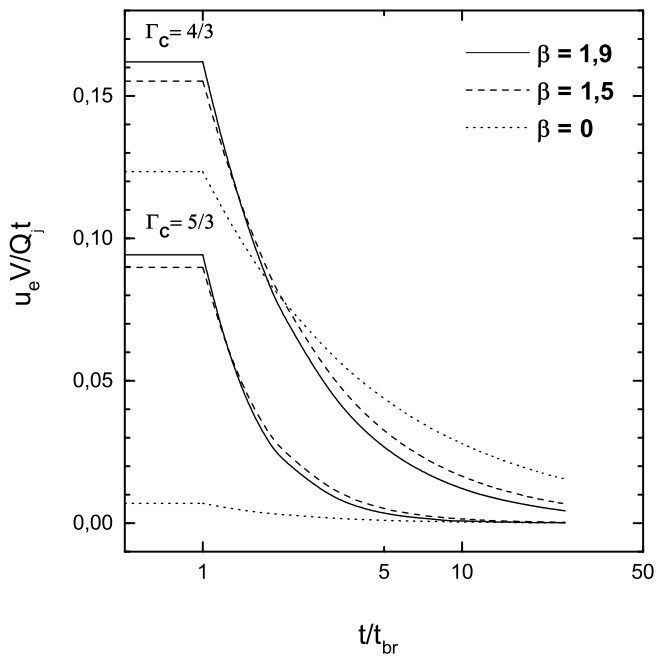}
\caption{Total energy of expanding lobes divided by jet power vs. time, calculated for three values of
$\beta$, for two general cases: $t/t_{\rm br}<$1 (KDA model) 
and $t/t_{\rm br}>$1 (KDA\,EXT model). All the
upper curves for a given value of $\beta$ are calculated for 
$\Gamma_{c}$=4/3, while the bottom lines for $\Gamma_{c}$=5/3. The other parameters are the same as in Fig.\,1.}
\label{7}
\end{figure}

\begin{figure}
\includegraphics[width=10cm, height=10cm]{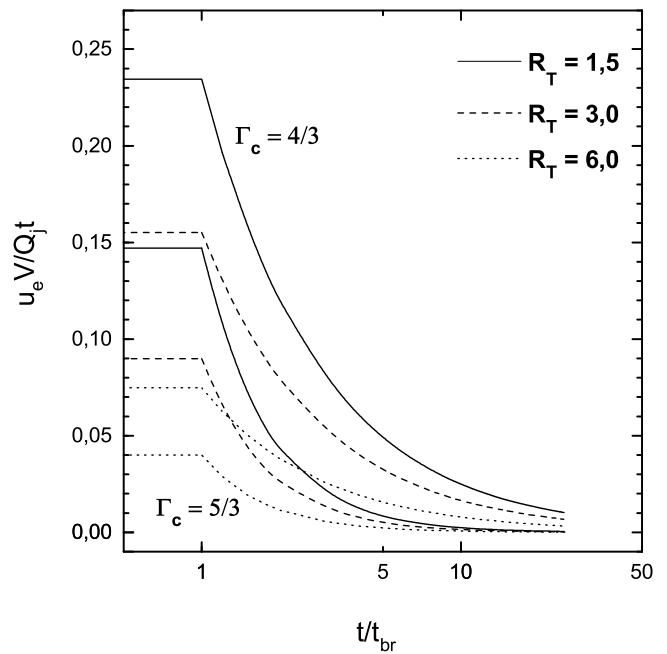}
\caption{As in Fig.\,7, but for three different values of $R_{T}$.}
\label{8}
\end{figure}

\subsection{Resulting radio spectra}

Though the $P_{\rm \nu}-D$ diagrams are very useful in diagnostics of the dynamical evolution of FR\,II$-$type radio sources, 
they suffer from one severe problem - the inability to verify their predictions by direct observations owing to 
the large timescale of the analysed processes. However, 
the analytical models of the evolution (including the KDA and KDA\,EXT) allow the source's 
luminosity to be predicted at a number of observing frequencies, i.e. its 
 radio spectrum. Here a few examples of such spectra, calculated for 
different sets of the fiducial source parameters, are shown and compared to the corresponding spectra 
calculated with the classical KDA model.

Figure\,9 presents the radio power spectra ($P_{\rm \nu}$ vs. $\nu$) of the fiducial source predicted with the 
KDA\,EXT model calculated for two different values of $\alpha_{\rm inj}$, and of two different values of 
jet switch-off time, $t_{\rm br}$. 
It is easy to see that their high-frequency slopes are close to the theoretical 
synchrotron aging spectrum of Jaffe \& Perola (1973); (cf. figure\,1 in Carilli et al. 1991). 

Figure\,10 shows the radio power spectra ($P_{\rm \nu}$ vs. $\nu$) of the fiducial source expected from the
KDA\,EXT model, calculated for two different values of redshift, $z$, as well as two different values of 
jet switch-off time, $t_{\rm br}$. 
It is worth noting that at the high redshift $z$=3, the decrease in radio power is much faster, 
which is consistent with the cosmological IGM models.

Figures\,11 and 12 show radio power spectra ($P_{\rm \nu}$ vs. $\nu$) of the fiducial source expected from the
KDA\,EXT model and calculated for two different values of $\beta$ and $R_{t}$, respectively. 
In these cases the spectra are also shown for two values of $t_{\rm br}$. Additionally, the 
distinction for $Q_{\rm j}$ values is introduced in order to ensure that the resulting plots will not be superimposed 
on each other (because the differences between the spectra of sources with various $\beta$ and 
$R_{t}$ are relatively low). The general trend indicates that the higher are $Q_{\rm j}$, $\beta$, and $R_{t}$, 
the higher are the values of the radio power, and the spectra resulting from the KDA\,EXT model are less curved at higher frequencies. 

The presented spectra indicate the very strong effect of the jet termination on both linear size and radio
luminosity of the lobes, especially for the values of $t_{\rm br}$ much lower than the
actual age of a source, $t$. The characteristic high-frequency break is 
more rapid in the case of low values of $\beta$ and $\rho_{0}$.
The strongest breaking of the spectra due to the rapid decrease in energy also occurs at the high 
redshift. All the presented spectra show how quickly the lobes of the FR\,II$-$type radio source may become 
invisible (or at least not in the detection range of contemporary radio telescopes) 
after the time $t=t_{\rm br}$, only if in the meantime the activity of the nucleus (AGN) has not restarted. 

\clearpage

\begin{figure}
\includegraphics[width=10cm, height=10cm]{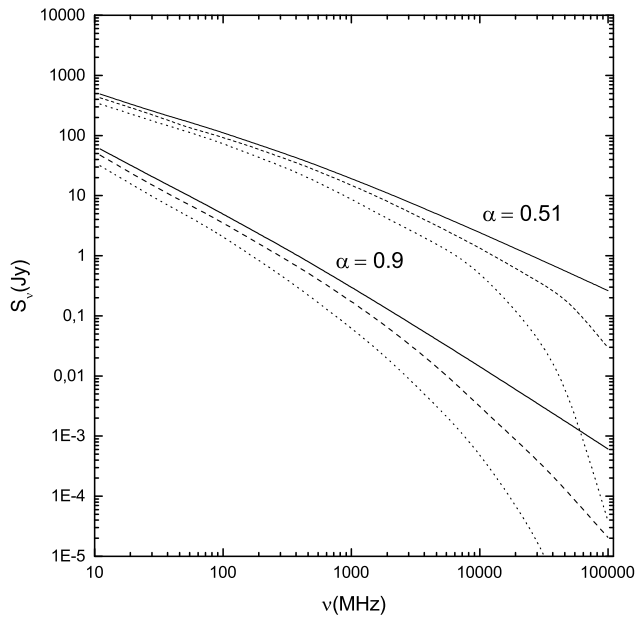}
\caption{Comparison of the radio power spectra ($P_{\nu}$ vs. $\nu$)
calculated with the KDA\,EXT model (dashed and dotted lines for $t_{\rm br}$=33 and 40 Myr, respectively) and the
KDA model (solid lines) for the fiducial source
with different $\alpha_{\rm inj}$.}
\label{9}
\end{figure}

\begin{figure}
\includegraphics[width=10cm, height=10cm]{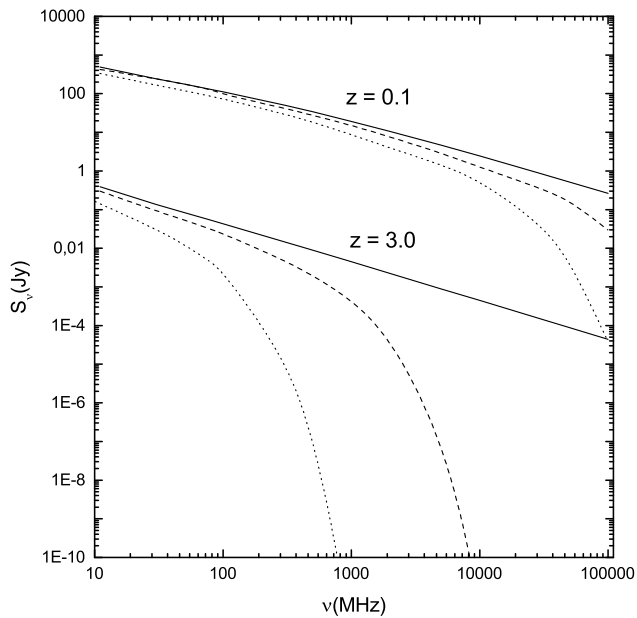}
\caption{Comparison of the radio power spectra ($P_{\nu}$ vs. $\nu$)
calculated with the KDA\,EXT model (dashed and dotted lines for $t_{\rm br}$=33 and 40 Myr, respectively) and the KDA model
(solid lines) for the fiducial source with different $z$. We note that the vertical scale is more extended than 
the other presented spectra in order to show the full shapes of the lower plots for $z$=3. }
\label{10}
\end{figure}

\begin{figure}
\includegraphics[width=10cm, height=10cm]{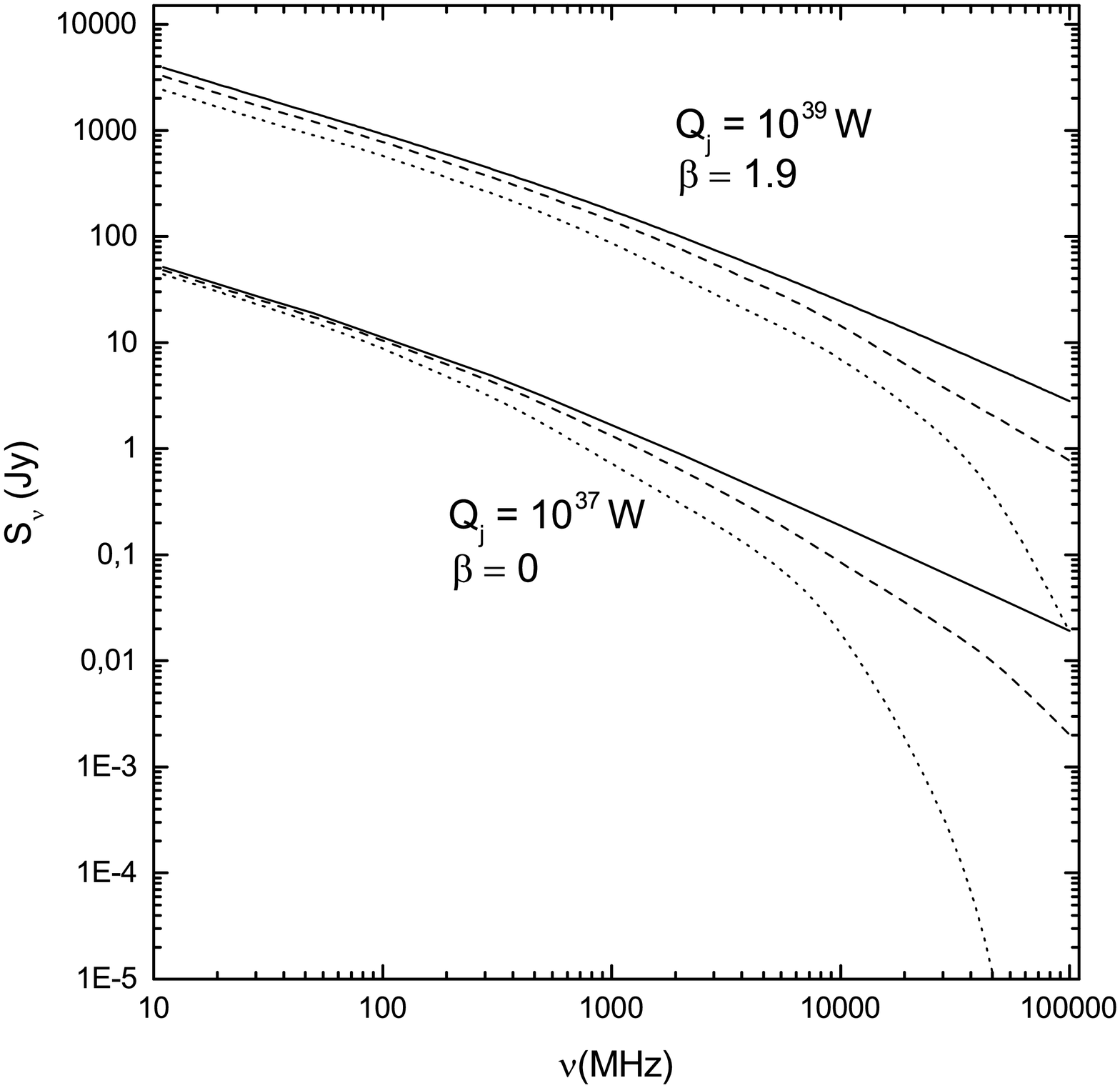}
\caption{Comparison of the radio power spectra ($P_{\nu}$ vs. $\nu$)
calculated with the KDA\,EXT model (dashed and dotted lines for $t_{\rm br}$ = 33 and 40 Myr, respectively) and the KDA model
(solid lines) for the fiducial source with different $\beta$ and $Q_{\rm j}$.}
\label{11}
\end{figure}

\begin{figure}[h]
\includegraphics[width=10cm, height=10cm]{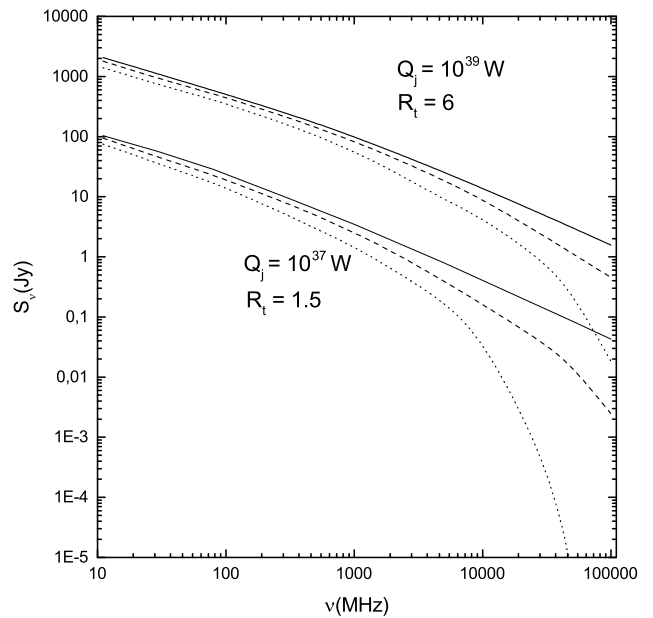}
\caption{Comparison of the radio power spectra ($P_{\nu}$ vs. $\nu$)
calculated with the KDA\,EXT model (dashed and dotted lines for $t_{\rm br}$ = 33 and 40 Myr, respectively) and the KDA model
(solid lines) for the fiducial source with different $R_{t}$ and $Q_{\rm j}$.}
\label{12}
\end{figure}

\clearpage

\section{Application of the KDA\,EXT model}
~
The KDA\,EXT model is expected to predict parameters of 
FR\,II$-$type radio sources with the terminated jet activity and ambient medium conditions for 
which the KDA model does not give reliable solutions. 
However, the spectra, sizes, and volumes of real sources are 
known from observations, and the values of the unknown 
'free' model parameters are the subject of our interest. The original KDA model, which assumes 
values for a number of its free parameters, enables the prediction of the time evolution of the
source's length $D$, and its radio luminosity $P_{\nu}$, at a given observing frequency. 
To solve the reverse problem of 
determining the values of $t$, $\alpha_{\rm inj}$, $Q_{\rm jet}$, and $\rho_{0}$ 
for a real radio source, it is possible to use the DYNAGE algorithm (Machalski et al. 2007) to find these values 
by a fit to its four observational parameters: size $D$, volume $V$, radio luminosity $P_{\nu}$ and 
radio spectrum $\alpha_{\nu}$, which provides $P_{\nu,i}$ at a number of observing frequencies $i$=1, 2, 3,...
This implies that it is also possible to verify the accuracy of the KDA\,EXT model for a given source with a highly steepened spectrum by deriving its spectra 
calculated from both the 'best age-solution' of the KDA model (DYNAGE) and a similar solution found with the KDA\,EXT model, and then comparing 
the results of these two models with the multifrequency observations of this source.

\subsection{Sample of examined radio sources}
~
Both models have been applied to six radio sources showing a strong bend in their 
observed radio spectra. This small sample includes three well-known 3C sources with small linear sizes and 
(in two cases) relatively high redshift (see Table~1). 
However, the KDA\,EXT model is expected to be especially useful in the case of giant FR\,II 
sources (i.e. J1428+3938; Machalski et al. 2006; Machalski 2011), which are 
also examined here. Two radio sources with restarting activity, J1453+3308 (Konar et al. 2006; 
Machalski et al. 2009) and J1548$-$3216 (Machalski et al. 2010), are  
included in the sample as examples of typical DDRG sources for which previous authors studied 
and modelled their individual lobes. Given the fact that observations of their secondary structure indicates 
the 'young' (renewed) jet emission, interacting and propagating within the previous 'old' emission in the same
 direction, it is interesting to examine how this may affect the 'old' primary structure (in particular its dynamical age).

Table~1 presents the observational data of the sources used for the fits.  
The errors in the flux measurements are taken from the references given under Table~1; however, 
in some cases great differences can be seen in the error values given in the data 
bases of the original publications. 
Moreover, some published errors are evidently underestimated, thus the relevant errors are 
arbitrariry enlarged, for example the errors claimed by Vigotti et al. (1999).

The 'largest angular size' ($LAS$) estimated from radio maps is given in arc seconds. The $R_{t}$ values are adopted from radio 
maps or the earlier publications. Table~1 also includes
the presumed inclination angle of the source's jet axis, $\theta$. The projected linear size of the sources, $D$, 
is calculated as

\begin{eqnarray}
D=4.848\times10^{-6}L_{\rm A}(z)LAS
\end{eqnarray}

\begin{center}
\begin{table*}[t]
\caption{\small Observational data for the selected FR II$-$type sources with strongly curved spectra.}
\label{table:1}
\footnotesize
\begin{tabular}{@{}lllllll}
\hline \hline
Name & 3C184 & 3C217 & 3C438 & J1428$+$3938 & J1453+3308 & J1548$-$3216  \\
\hline
$z$ & 0.994 & 0.898 & 0.290 & 0.5 (est)  & 0.249 & 0.1082 \\
$LAS$ [\arcsec] & 4.8 & 12.1 & 22.4 & 269 &  336 & 522 \\
$R_{t}$ & 3.0 & 4.0 & 2.7 & 3.4 & 3.8 & 2.9 \\
$D$ [kpc] & 38.5 & 94 & 100 & 1630  & 1297 & 998 \\ 
$\theta$ [$^{o}$] & 90 & 70 & 90 & 90 & 90 & 90 \\
\hline
& & & & & &\\
$\nu$ [MHz] & & & $S_{\nu}\pm\Delta S_{\nu} $ [mJy] &\\
& & & & & & \\
22 & 61000$\pm$5000 [14] & & & & & \\
26 & & 43000$\pm$6000 [8] & 228000$\pm$12000 [8]  & & &\\
38 & 37200$\pm$3720 [7] & 39000$\pm$5850 [9] & 150290$\pm$40650 [19] & & &\\
38 & &  & 162840$\pm$48850 [8] & & &\\
74 & 23660$\pm$2370 [10] & 25380$\pm$2590 [10] & 81560$\pm$8160 [10] & & &\\
86 & 23800$\pm$2700 [8] & 24400$\pm$1300 [8] & 86700$\pm$8720 [8] & & &\\
151 & 14870$\pm$893 [7] & 16498$\pm$294 [21] & & 971$\pm$109 [4] & 2165$\pm$110  &\\
151 & & & & 990$\pm$95 \,\,\,[7] &   &\\
160 & & & & & & 8400$\pm$840  \\
178 & 14652$\pm$700 [9] & 15590$\pm$600 [20] & 51400$\pm$4100 [19] &  &  2020$\pm$200  &  \\
178 & 14388$\pm$719 [8] &  & 49620$\pm$2300 [19] &  &   & \\
232 & & & & 810$\pm$100 [3] & &  \\
240 & & & & & 1667$\pm$250  &  \\
325 & 9676$\pm$350 [1] & 8970$\pm$356 [1] & 29940$\pm$1080 [1] & 428$\pm$34 \,\,\,[1] &  1365$\pm$140  &  \\
334 & & & & & 1456$\pm$112  & 4737$\pm$710  \\
365 & 9083$\pm$328 [11] & 8372$\pm$242 [11] & 26400$\pm$950 [11] & & &\\
408 & & 7840$\pm$610 [20] & 23760$\pm$480 [15] & & & \\
408 & & 7090$\pm$280 [6] & 25180$\pm$1980 [19]   & 270$\pm$31 \,\,\,[6]  &  &  \\
605 & & & & & 970$\pm$75   &  \\
619 & & & & & & 3141$\pm$252  \\
750 & 4230$\pm$177 [8] & 4060$\pm$208 [8] & 13700$\pm$700 [19] & & &\\
750 & 4200$\pm$200 [9] & 4300$\pm$140 [20] & & & &\\
1287 & & & & & 442$\pm$34   & \\
1384 & & & & & &  1733$\pm$87 \\
1400 & 2582$\pm$78 [2] & 2087$\pm$63 [2] & 6854$\pm$220 [2] & 83$\pm$4\,\,\,\,\,\,\,\,\,\,[2]  & 426$\pm$36  & \\
1400 & 2275$\pm$170 [18] & 2230$\pm$70 [20] & 6940$\pm$220 [19] &  &   & \\
1477 & & & 6410$\pm$128 [16] & & & \\
2495 & & & &  &  & 963$\pm$30 \\
2695 & 1183$\pm$60 [8] & 1011$\pm$80 [20] & 3260$\pm$155 [19] & & &\\
2700 & & 1020$\pm$50 [20]  & 3300$\pm$160 [19] & & &\\
4830 & &  & 1635$\pm$164 [17] & & &\\
4850 & 618$\pm$455 [12] & 550$\pm$49 [12] & 1607$\pm$143 [18] & & &\\
4860 & & & & 13$\pm$3\,\,\,\,\,\,\,\,\,\,[5]  & 104$\pm$8  & 415$\pm$42\\
4900 & &  & 1580$\pm$60 [19] & & &\\
5000 & 596$\pm$40 [20] & 477$\pm$60 [8] & 1529$\pm$60 [19] & & & \\ 
8440 & &  & 765$\pm$38 [13] & & &\\
10550 & &  & &  3$\pm$2\,\,\,\,\,\,\,\,\,\,\,\,\,[6] &  &  \\
10695 & & 122$\pm$38 [8] & 640$\pm$40 [19] & & & \\
10705 & 216$\pm$30 [8] & 130$\pm$40 [9] & 600$\pm$38 [8] & & &\\
14900 & 150$\pm$20 [8] & 120$\pm$20 [8] & 390$\pm$30 [8] & & &\\
\hline
\end{tabular}
\tablebib{
[1] {WENSS (Rengelink et al., 1997)}; [2] {NVSS (Condon et al., 1998)}; [3] {Miyun (Zhang et al., 1997)}; [4] {7C (Waldram et al., 1996)};
[5] {(Machalski et al., 2006)}; [6] {B3-VLA (Vigotti et al., 1999)}; [7] {6CII (Hales et al. 1988)}; [8] {(Laing \& Peacock 1980)}; 
[9] {Kellermann \& Pauliny-Toth 1973)}; [10] {VLSS (Cohen et al., 2007)}; [11] {TXS (Douglas et al., 1996)}; [12] 
{GB6 (Gregory et al., 1996)}; [13] {(Hardcastle at el. 1998)}; [14] {(Roger et al. 1986)}; [15] {B3 (Ficarra et al. 1985)}; 
[16] {(Leahy \& Pearley 1991)}; [17] {Griffith et al. 1991}; [18] {(White \& Becker 1992)}; [19] {(K\"{u}hr et al. 1981)}; 
[20] {K\"{u}hr et al. 1979}; [21] {7CN (Riley et al. 1999)}. The flux densities for 
J1453+3308 and J1548$-$3216 are adopted from Konar et al. (2006) and Machalski et al. (2010), respectively.
}
\end{table*}
\end{center}

The observed flux densities for the 3C sources are taken from the NED NASA database. 
The flux densities for J1428$+$3938 are derived from various radio surveys
(7C, B3-VLA, WENSS, NVSS, Miyun). The flux density data from dedicated observations
 with the GMRT and VLA arrays for the outer lobes of J1453$+$3308 and J1548$-$3216 are
taken from the original publications in order to model their extended, outer FR\,II$-$type structures only.
Where it was possible, original flux densities were previously corrected by substacting the core emission from 
the total flux of the sources. The values of the sources' radio power $P_{\nu}$ - which are the proper physical 
observables to use in the modelling - are calculated from the given flux densities according to the formula  

\begin{eqnarray}
P_{\nu}=S_{\nu} L_{\rm D}(z)^{2} (1+z)^{(\alpha_{\nu}-1)}
\end{eqnarray}

\noindent
where $L_{\rm A}(z)$ (in Equation\,13) and $L_{\rm D}(z)$ (in Equation\,14) are the angular and luminosity 
distances of the source, respectively, determined with the Cosmological Calculator of 
Wright (2006) assuming a flat Universe with Hubble constant $H_{0}$=71 ${\rm km \, s^{-1}Mpc^{-1}}$ and the $\Lambda$CDM model with cosmological parameters 
$\Omega_{m}$=0.27 and $\Omega _{\Lambda}$=0.73, and where $\alpha_{\nu}$ is the spectral index measured as the 
spectrum slope gradient calculated separately for every pair of neighbouring radio flux densities.

~It is worth noting a need to know the source's radio power $P_{\nu}$ for at least five observing
frequencies for a proper application of the KDA and KDA\,EXT models. In general, the wider the 
frequency coverage of the radio spectra of the source, the more precise is the model fit. 
This indicates that it is also advisable to know the observed flux densities for the very low 
(about 10 - 20 MHz) and very high (above 10 GHz) radio frequency in order to take it 
into account in the modelling. 
However, 18-25 flux density data points covering the frequency range 22 MHz to 15 GHz are 
provided for the 3C sample sources. For the three remaining sources presented in Table~1 there 
are at least seven flux densities covering the quite broad frequency range, although some hopes 
are connected with the new LOFAR interferometer.

\subsection{Fitting procedure}

The fitting method consists of two steps. In the first, the best 
'age-solution' of the KDA model (t$_{\rm KDA EXT}$; Table~8) is determined using the 
DYNAGE algorithm. However, as expected, the best model fit 
and resulting spectra remain unsatisfied as long as we are trying to reproduce observed 
highly steepened spectrum with the model assuming the CI process of the energy supply to the lobes 
(the case of the KDA model). To solve this problem, I performed the KDA fits using only 
the low-frequency parts of the flux density data, i.e. rejecting the extremely high-frequency fluxes 
that increase the curvature of the source's observed spectra.  

Following the original KDA analysis, their Case 3 where both the cocoon material and the 
ambient medium are assumed to be described by the 'cold' equation of state 
(i.e. $\Gamma_{\rm c}$=$\Gamma_{\rm a}$=$\Gamma_{\rm B}$=$5/3$) is adopted.
The equipartition condition for the initial ratio of the energy densities between
 the source's magnetic field and relativistic particles, $r \equiv u_{\rm B}/u_{\rm e}$=$(1+p)/4$,
 is also assumed, which is well supported by the X-ray observations of the lobes in powerful radio sources 
(Kataoka \& Stawarz 2005). Finally, following Daly (1995) and Blundell et al. (1999), 
I adopt $\beta$=1.5 for all of the examined sources. The remaining model free parameters are the
same as for the fiducial source in Section\,3. The final result of this step is the determination of four model parameters: the initial 
particle-energy distribution described by $\alpha_{inj}$; the fitted age, $t$; the corresponding
 values of the jet power, $Q_{j}$; and the central core density, $\rho_{0}$.

In the second step, a number of values for the jet switch-off time $t_{br}$, 
fulfilling the $t_{br}<t$ condition, is selected. For each of these values, Equations.\,8, i.e. 
Equations\,3 and 9 are solved numerically, providing the source power, $P_{\nu_{em}}$, at the number of frequencies
(in the source frame) corresponding to a given $t_{br}$ value. The best fit is determined using
the least-squares method by minimizing the expression

\begin{eqnarray}
\chi^{2}_{red}=\frac{1}{n-3}\sum\limits_{n}\left(\frac{S_{\nu_{0}}-S_{MOD}}{\Delta S_{\nu_{0}}}\right)^2,
\end{eqnarray}  
\noindent
where $S_{\nu_{0}}$ and $\Delta S_{\nu_{0}}$ are flux densities 
and $S_{MOD}$ are the model flux densities re-calculated from the model
values of $P_{\nu_{em}}$ according to 

\begin{eqnarray}
S_{\rm MOD}=P_{\nu_{em}} \left(\frac{1+z}{L_{D}^{2}(z)}\right) = P_{\nu_{0}(1+z)} \left(\frac{1+z}{L_{D}^{2}(z)}\right).
\end{eqnarray}

The $S_{\rm MOD}$ flux densities, resulting from the KDA and
KDA EXT models, are given in Tables~2 - 7 and are compared to the
observed flux densities which determine radio spectra of the
sample sources.

\subsection{Fitting results}

~
In order to compare the KDA\,EXT fits with the KDA solutions, 
and with similar results already published for the
 sample sources (including those from the KDA model for entire sources or their radio lobes only),
 the values of $\chi^{2}_{red}$, determining the goodness of fit, are calculated for
 all of the best fits of the KDA models and the best fit of the KDA\,EXT model.
The latter fits are characterized by $\alpha_{inj}$ values that are flatter and dynamical ages that are 
greater (especially for larger and older sources) than those with the KDA fits.

Table~8 presents values of the model free parameters and some derivative physical 
parameters of the sample sources derived from their best fit of the KDA and KDA\,EXT models. These 
derivative parameters are 
the cocoon pressure, $p_{\rm c}$; the total emitted energy, 
U$_{\rm c}$; the strength of the magnetic field, B$_{\rm eq}$; and the radial expansion speed 
of the cocoon's head, v$_{\rm h}$. It is worth noting that their values are calculated 
for t $>$ t$_{br}$ (after the switch-off). Namely, $p_{\rm c}$ for t>t$_{br}$ is calculated 
with Equation 12, U$_{\rm c}$ is given by the formula 
U$_{\rm c}(t)$=u$_{\rm c}(t)$V$_{\rm c}(t)$=p$_{\rm c}(t)$V$_{\rm c}(t)/(\Gamma_{c}-1)$, 
and v$_{\rm h}(t)$ with Equation 11. A compilation of the results obtained for the individual 
sources from the sample is presented below. 

\subsubsection{3C184}
~3C184 is an example of powerful high-redshift radio galaxies embedded in a 
cluster environment (Belsole et al. 2004). Its lobe length is rather small and equal to ~38.5 
kpc. It may then be supposed that this source is relatively young. 
It is worth noting that in the case of 3C184 (and in two other 3C sources examined in 
this paper) its radio spectrum is observed within a wide range of frequencies extending from 
22/26 {\rm MHz} to even 15 {\rm GHz}.

The spectra resulting from the predicted flux densities $S_{\rm MOD}\,(KDA)$ and 
$S_{\rm MOD}\,(KDA\,EXT)$ are shown in Figure 13. The KDA\,EXT spectrum is evidently much 
more bent than the KDA spectrum and fits better to the flux densities measured at both, low and high 
frequencies. The goodness of the fit to data points ($\chi^{2}_{red}$ values at the bottom of 
Table~2) clearly indicate a superiority of the KDA\,EXT solution over the KDA.

3C184 is the only source in my sample whose fitted age is less than 1 Myr. However, this 
young age is accompanied by low relative expansion speed of the lobes' head $v_{\rm h}/c$, 
being only about 2-3 times higher than these for the three giant-sized sources in the sample. 
This is concordent with the 'best-fit' age solution with the assumed jet termination. Perhaps 
it suggest that this high redshift structure is not at an early stage of its evolution, and 
will not evolve into a structure much larger than 100 kpc.  

\begin{table}[here]
\caption{\small Flux densities resulting from the KDA and KDA\,EXT models for 3C184 and their goodness of fit to the observed data.}
\footnotesize
\begin{center}
\begin{tabular}{@{}rrrr}
\hline
$\nu_{0}$ & $S_{\nu_{0}}\pm\Delta S_{\nu_{0}}$ & $S_{MOD}$  & $S_{MOD}$ \\
         & & KDA  & KDA EXT\\
\hline
\hline
22 & \,\,\,61000\,$\pm$\,5000 & 69177 & 60377 \\
38 & \,\,\,37200\,$\pm$\,3720 & 46877 & 42388 \\
74 & \,\,\,23660\,$\pm$\,2370 & 28377 & 26961 \\
86 & \,\,\,23800\,$\pm$\,2700& 25407 & 24262 \\
151 & 14870\,$\pm$\,893 & 16179 & 16116 \\
178 & 14652\,$\pm$\,700 & 14124 & 14240 \\
178 & 14388\,$\pm$\,719 & 14124 & 14240 \\
325 & \,\,\,9676\,$\pm$\,350 & 8472 & 8887 \\
365 & \,\,\,9083\,$\pm$\,328 & 7655 & 8088 \\
750 & \,\,\,4230\,$\pm$\,177 & 4018 & 4377 \\
750 & \,\,\,4200\,$\pm$\,200 & 4018 & 4377 \\
1400 & 2582\,$\pm$\,78 & 2244 & 2456 \\
1400 & 2275\,$\pm$\,170 & 2244 & 2456 \\
2695 & 1183\,$\pm$\,60 & 1191 & 1254 \\
4850 & \,\,\,618\,$\pm$\,55 & 665 & 643 \\
5000 & \,\,\,596\,$\pm$\,40 & 646 & 621 \\
10705 & \,\,\,216\,$\pm$\,30 & 299 & 238 \\
14900 & \,\,\,150\,$\pm$\,30 & 212 & 146 \\
\hline
$\chi^{2}_{red}$ &  & 5.94 & 1.89 \\
\hline
\end{tabular}
\end{center}
\end{table}

\begin{figure}[here]
\includegraphics[width=10cm, height=10cm]{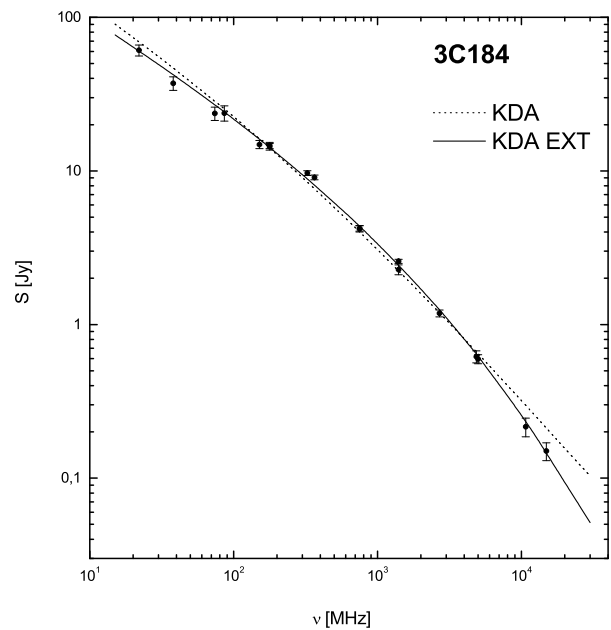}
\caption{Best KDA fit (dotted line) and KDA\,EXT fit (solid line) for the radio galaxy 3C184. The values of flux density and frequency are presented in logarithmic scale. 
The observed flux densities and their errors are marked with data points.}
\label{13}
\end{figure}

\subsubsection{3C217}
~
This high-redshift radio galaxy was included in Sample 3 (3CRR FR\,II$-$type sources with z>0.5 and D<400 {\rm kpc}) in 
Kuligowska et al. (2009). 3C217 is another example of a relatively young and small 
(D=94 {\rm kpc}) source with a strongly bent radio spectrum. 

The spectra resulting from the KDA and KDA\,EXT models and overlain the data points are shown in 
Figure\,14. Similarly to 3C184, the KDA\,EXT model spectrum is much more bent than that predicted 
by the KDA model. The $\chi^{2}_{red}$ values at the bottom of Table~3 clearly indicate a 
significantly higher goodness of fit in the case of the KDA\,EXT solution.  

\begin{table}[here]
\caption{\small Flux densities resulting from the KDA and KDA\,EXT models for 3C217 and their goodness of fit to the observed data.}
\footnotesize
\begin{center}
\begin{tabular}{@{}rrrr}
\hline
$\nu_{0}$ & $S_{\nu_{0}}\pm\Delta S_{\nu_{0}}$ & $S_{MOD}$  & $S_{MOD}$ \\
        & & KDA  & KDA EXT\\
\hline
\hline
26 & \,\,\,43000\,$\pm$\,6000 & 57885 & 53440 \\
38 & \,\,\,39000\,$\pm$\,5850 & 44624 & 42332 \\  
74 & \,\,\,25380\,$\pm$\,2590& 27200 & 27396 \\
86 & \,\,\,24400\,$\pm$\,1300& 24242 & 24688 \\
151 & 16498\,$\pm$\,294 & 15544 & 16227 \\
178 & 15590\,$\pm$\,600 & 13601 & 14287 \\
325 & \,\,\,8970\,$\pm$\,324 & 8214 & 8989 \\
365 & \,\,\,8372\,$\pm$\,242 & 7440 & 8168 \\
408 & \,\,\,7090\,$\pm$\,280 & 6754 & 7398 \\
408 & \,\,\,7840\,$\pm$\,610 & 6754 & 7398 \\
750 & \,\,\,4060\,$\pm$\,208 & 3923 & 4127 \\
1400 & 2086\,$\pm$\,62 & 2201 & 2126 \\
1400 & 2230\,$\pm$\,70 & 2201 & 2126 \\
2695 & 1011\,$\pm$\,80 & 1171 & 1038 \\
2700 & 1020\,$\pm$\,50 & 1168 & 1038 \\
4850 & \,\,\,550\,$\pm$\,49 & 654 & 509 \\
5000 & \,\,\,477\,$\pm$\,60 & 635 & 487 \\ 
10695 & \,\,\,122\,$\pm$\,38 & 294 & 150 \\
10705 & \,\,\,130\,$\pm$\,40 & 293 & 149 \\ 
14900 & \,\,\,120\,$\pm$\,20 & 201 & 84 \\ 
\hline
$\chi^{2}_{red}$ &  & 8.00 & 1.21 \\
\hline
\end{tabular}
\end{center}
\end{table}

\begin{figure}
\includegraphics[width=10cm, height=10cm]{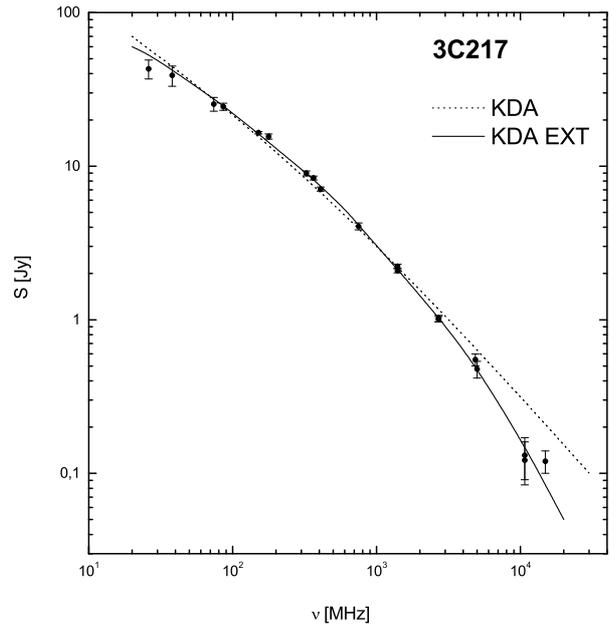}
\caption{
As in Fig.\,13, but for 3C217.
}
\label{14}
\end{figure}

\subsubsection{3C438}
~
Similarly to 3C217, 3C438 is a typical but low-redshift FR\,II$-$type source with linear size 
not exceeding 100 {\rm Mpc}. Also, its observed radio spectrum shows considerable bending from 
the middle frequencies to the higher ones.
The spectra predicted with the KDA and KDA\,EXT models are shown in Figure 15. Again, the relevant 
$\chi^{2}_{red}$ values in these two models are similar to those for 3C184 and 3C217, and confirm 
the superiority of the KDA\,EXT age solution.
 
\begin{table}[here]
\caption{\small Flux densities resulting from the KDA and KDA\,EXT models for 3C438 and their goodness of fit to the observed data.}
\footnotesize
\begin{center}
\begin{tabular}{@{}rrrr}
\hline
$\nu_{0}$ & $S_{\nu_{0}}\pm\Delta S_{\nu_{0}}$ & $S_{MOD}$  & $S_{MOD}$ \\
         & & KDA  & KDA EXT\\
\hline
\hline
26 & \,\,\,228000\,$\pm$\,12000 & 229220 & 243060 \\
38 & \,\,\,150290\,$\pm$\,40650 & 171455 & 182345 \\
38 & \,\,\,16284\,$\pm$\,48850 & 171455 & 182345 \\
74 & \,\,\,81560\,$\pm$\,8160 & 99095 & 105770 \\
86 & \,\,\,86700\,$\pm$\,8720 & 87287 & 93230 \\
178 & 51400\,$\pm$\,4100 & 46105 & 49377 \\
178 & 49620\,$\pm$\,2300 & 46105 & 49377 \\
325 & \,\,\,29940\,$\pm$\,1080 & 26506 & 28440 \\
365 & \,\,\,26400\,$\pm$\,950 & 23766 & 25505 \\
408 & \,\,\,23760\,$\pm$\,480 & 21387 & 22959 \\
408 & \,\,\,25180\,$\pm$\,1980 & 21387 & 22959 \\
750 & \,\,\,13700\,$\pm$\,700 & 11876 & 12727 \\
1400 & 6940\,$\pm$\,220 & 6378 & 6755 \\
1400 & 6854\,$\pm$\,220 & 6378 & 6755 \\
1477 & 6410\,$\pm$\,128 & 6042 & 6388 \\
2695 & 3260\,$\pm$\,155 & 3260 & 3319 \\
2700 & 3300\,$\pm$\,160 & 3259 & 3311 \\
4830 & 1635\,$\pm$\,164 & 1779 & 1638 \\   
4850 & \,\,\,1607\,$\pm$\,143 & 1771 & 1629 \\
4900 & 1580\,$\pm$\,60 & 1752 & 1607 \\
5000 & 1529\,$\pm$\,60 & 1752 & 1565 \\
8440 & \,\,\,765\,$\pm$\,38 & 986 & 825 \\
10695 & \,\,\,640\,$\pm$\,40 & 766 & 619 \\
10705 & \,\,\,600\,$\pm$\,38 & 766 & 619 \\
14900 & \,\,\,390\,$\pm$\,30 & 538 & 407 \\
\hline
$\chi^{2}_{red}$ &  & 8.57 & 1.17 \\
\hline
\end{tabular}
\end{center}
\end{table}

\begin{figure}[here]
\includegraphics[width=10cm, height=10cm]{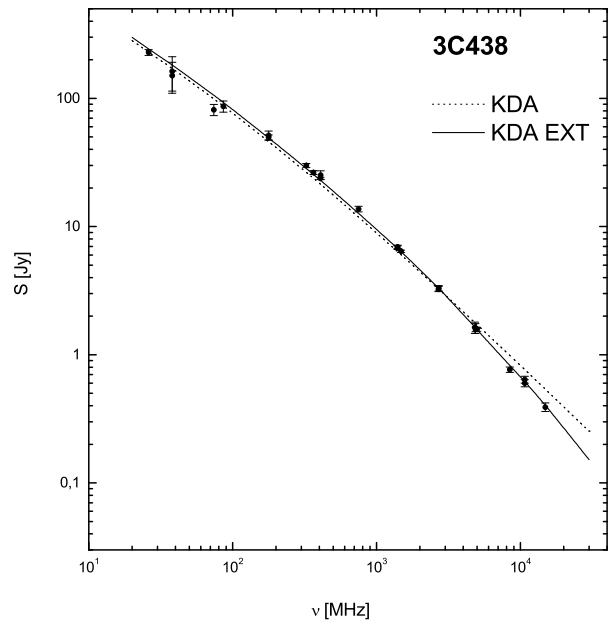}
\caption{As in Fig.\,13, but for 3C438.
}
\label{15}
\end{figure}

\subsubsection{J1428$+$3938}
~
J1428$+$3938 is a giant-sized (D $>$ 1 {\rm Mpc}) radio galaxy identified as a giant radio 
galaxy (GRG) by Machalski et al. (2006). 
Its central radio core that is precisely coincident with a faint optical galaxy with R=21.11 mag implies a 
high redshift; unfortunately, it has not been confirmed spectroscopically.
Therefore, the authors estimate its value from the Hubble $m_{R}-z$ relation for giant radio sources as about 0.5.
Extended radio lobes of this source have been modelled by Machalski (2011) using the DYNAGE algorithm -- contrary 
to a simpler approach where a dynamical model is fitted to
the spectrum of entire source, i.e. to a mean of its two lobes. This is worth emphasizing that such an approach 
is only possible when a lack of angular resolution
does not allow for discrimination between spectra of opposite lobes, especially at low observing frequencies.

In order to compare the best determined KDA and KDA\,EXT model solution to the fit resulting from the Machalski 
model for the lobes, I use the arbitrary means
of $\alpha_{\rm inj}$, $t$, $Q_{\rm jet}$ and $\rho_{0}$ values given in his Table~2. These mean values are as follows: $\alpha_{\rm inj}$=0.532, $t$=415 {\rm Myr},
$Q_{jet}$=3.93$\times$$10^{37}$ {\rm W} and $\rho_{0}$=1.46$\times$$10^{-22}$ kg/m$^{3}$. In addition to these 
values, the above model (hereafter M2011) assumed different
equations of state, $\Gamma_{c}$=$\Gamma_{B}$=4/3.

It is particularly interesting to examine how the KDA\,EXT model fits the observed spectrum of this source. I use the 151 {\rm MHz} measurements as the lowest one and
include the very uncertain Miyun measurement at 232 {\rm MHz}. The original radio flux density of 6.2 {\rm mJy} 
at 10550 {\rm MHz} taken from the B3-VLA catalogue (Vigotti et al.
1999), likely referring to the entire source, is corrected here by subtracting the probable flux density of the radio core, supposed to dominate the total radio flux at the high frequencies.
This is justified by the 4860 {\rm MHz} VLA observations showing the core flux of 3.5 {\rm mJy}. 

The spectra resulting from the KDA and KDA\,EXT models and including the data points 
are shown in Figure\,16. The goodness of the fits indicate that the continuum-injection 
spectra predicted with the M2011 and KDA models are quite comparable, 
though the fitted values of their free parameters are significantly different.
This emphasizes the problem of how to determine the best age solution, reproducing the observed 
spectrum with a comparable accuracy (again see Brocksopp et al 2011).
The KDA\,EXT model fits the observed data better than KDA.
The ($t-t_{br}$)/$t_{br}$ ratio of about 0.1 (Table~8) strongly supports the 
hypothesis that the observed structure of the lobes and their spectrum reflect the
evolutionary phase after the termination of the jet activity.

According to Figure\,16, the flux density measurement at 232 \,MHz is a clear outlier from the 
plot and its measurement uncertainty is high. It is reflected in the relatively high 
values of $\chi^{2}_{red}$ for the results of modelling. It can be shown that removing this data 
point from the set reduces the resulting $\chi^{2}_{red}$ for the KDA\,EXT model to the value of
 1.44. However, in the case of the KDA solutions it gives the simultaneous increase of this 
value to about 18.  

The most interesting thing here is that if J1428$+$393 is really at the assumed redshift, it is the
 old and large source with the evidence of the strong aging in opposition to the two DDRG sources 
J1453$+$3308 and J1548$-$3216. Furthermore, owing its supposed high redshift, the inverse Compton 
losses should dominate over synchrotron losses. Indeed, once the jet activity ceases, a strong 
aging will occur.   

\begin{table}[here]
\caption{\small Flux densities resulting from the KDA and KDA\,EXT models for J1428$+$3938 and their goodness of fit to the observed data}
\footnotesize
\begin{center}
\begin{tabular}{@{}rrrrr}
\hline
$\nu_{0}$ & $S_{\nu_{0}}\pm\Delta S_{\nu_{0}}$ & $S_{MOD}$ & $S_{MOD}$ & $S_{MOD}$  \\
    & & M2011 & KDA & KDA\,EXT  \\
\hline
\hline
151 & \,\,\,971\,$\pm$\,109& 919  & 995 & 891  \\
151 & 990\,$\pm$\,95 & 919 & 995 & 891  \\
232 & \,\,\,810\,$\pm$\,100 & 602 & 617 & 582   \\
325 & 428\,$\pm$\,34 & 431 & 421 & 413   \\
408 & 270\,$\pm$\,31 & 344 & 324 & 326   \\
1400 & 83\,$\pm$\,4 & 99 & 73 & 83   \\
4860 & 13\,$\pm$\,3 & 28 & 15 & 15   \\
10550 & \,\,\,3\,$\pm$\,2 & 13 & 6 & 4   \\
\hline
$\chi^{2}_{red}$ &  & 15.52 & 3.69 & 2.20  \\
\hline
\end{tabular}
\end{center}
\end{table}

\begin{figure}
\includegraphics[width=10cm, height=10cm]{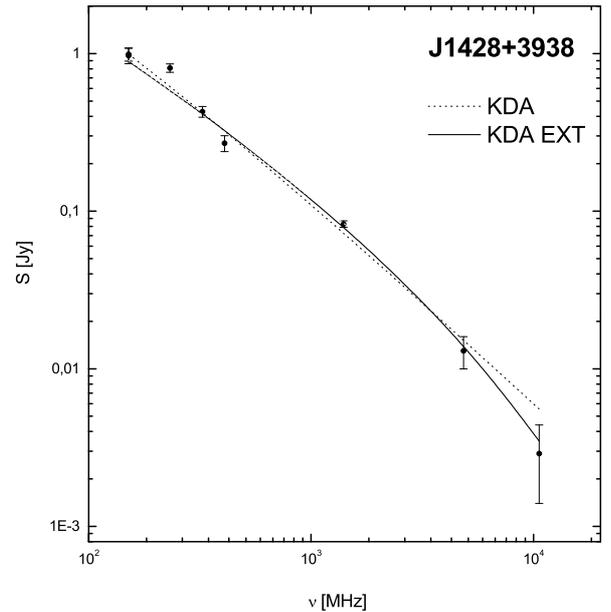}
\caption{As in Fig.\,13, but for J1428$+$3938.}
\label{16}
\end{figure}

\subsubsection{J1453$+$3308}
~
J1453$+$3308 is the well-known 
DDRG galaxy whose radio structure has already been studied by several authors 
(e.g. Schoenmakers et al. 2000; Kaiser et al. 2000; Konar et al. 2006). It consists of two 
pairs of lobes, the extended outer lobes and slimmer 
inner ones, considered to be the result of secondary nuclear activity occurring after a termination of the primary jets. 
Spectral aging analysis of these pairs of lobes was 
published by Konar et al. (2006), while their dynamical analysis was undertaken by Machalski et al. 
(2009) (in the frame of a subsample of ten GRGs), and, independently by Brocksopp et al. (2011).

Also in this case, the best-fit KDA and KDA\,EXT flux densities shown in Figure~17 models were 
compared to the observed data and to the flux densities of Machalski et al. (2009) model 
(hereafter MJS2009) for the outer lobes. The arbitrary mean values of the model parameters used for a comparison 
are $\alpha_{inj}$=0.547, $t$=146 {\rm Myr}, $Q_{jet}$=4.96$\times$$10^{37}$ {\rm W}, and 
$\rho_{0}$=2.51$\times$$10^{-23}$ kg/m$^{3}$. Additionally, the MJS2009 model assumes $\Gamma_{B}$=4/3. The 
resulting flux densities and the goodness of the fit for 
each model are given in Table~6. The values of $\chi^{2}_{red}$ indicate that the KDA\,EXT fit 
is better than MJS2009 and KDA, though the differences in the goodness of the fits are low. 
These two continuum injection models reproduce well the low-frequency part of the observed 
spectrum; starting at some point, the KDA\,EXT spectrum is clearly more bent towards high 
frequencies. Again, the model spectra predicted with the MJS2009 and KDA models are comparable in
 spite of the quite different age solutions (reflecting the problem mentioned earlier), and 
$\chi^{2}_{red}$ value in the KDA\,EXT model is similar to that for J1428$+$3938. It should be 
noticed that the models presented here concern the outer (primary) lobes of these DDRGs only.
 Because the new (secondary) young inner lobes are also observed, it is expected that 
these outer lobes have had no time for noticeable aging effects. 

\begin{table}[h]
\caption{\small Flux densities resulting from the KDA
 and KDA\,EXT models for J1453$+$3308 and their goodness of fit to the observed data.}
\footnotesize
\begin{center}
\begin{tabular}{@{}rrrrr}
\hline
$\nu_{0}$  & $S_{\nu_{0}}\pm\Delta S_{\nu_{0}}$ & $S_{MOD}$ & $S_{MOD}$ & $S_{MOD}$ \\
        & & MJS2009  & KDA & KDA\,EXT \\
\hline
\hline
151 & \,\,\,\,2165\,$\pm$\,110 & 2475 & 2391 & 2399 \\  
178 & \,\,\,\,2020\,$\pm$\,200 & 2179 & 2136 & 2158 \\
240 & \,\,\,\,1667\,$\pm$\,250 & 1715 & 1724 & 1751 \\ 
325 & \,\,\,\,1365\,$\pm$\,140 & 1332 & 1379 & 1378 \\ 
334 & \,\,\,\,1456\,$\pm$\,112 & 1301 & 1351 & 1347 \\
605 & \,\,\,\,970\,$\pm$\,75 & 771 & 847 & 843 \\
1287 & \,\,\,\,442\,$\pm$\,34 & 389 & 441 & 426 \\
1400 & \,\,\,\,426\,$\pm$\,26 & 352 & 410 & 394 \\
4850 & \,\,104\,$\pm$\,8 & 104 & 125 & 115 \\
\hline
$\chi^{2}_{red}$ & & 4.80 & 2.54 & 2.09 \\
\hline
\end{tabular}
\end{center}
\end{table}

\begin{figure}
\includegraphics[width=10cm, height=10cm]{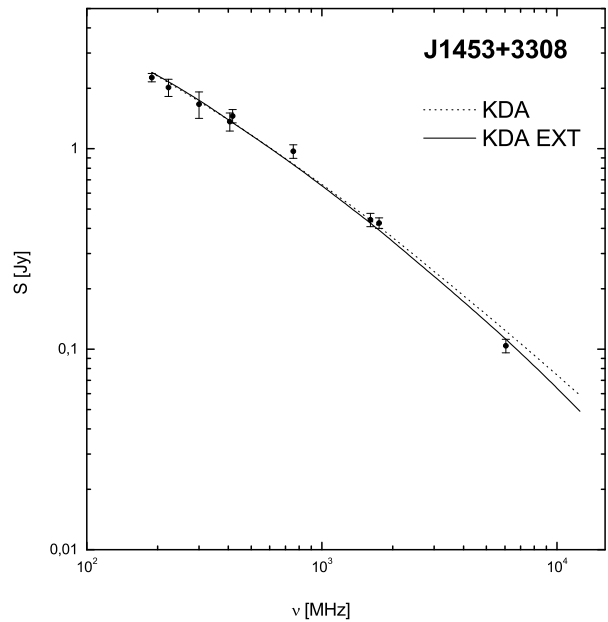}
\caption{As in Fig.\,13, but for J1453$+$3308.}
\label{17}
\end{figure}

\subsubsection{J1548$-$3216}
~
J1548$-$3216 (PKS B1545$-$321) is another example of a giant DDRG. This low-redshift radio galaxy 
was discovered by Saripalli et al. (2003). Its dynamical analysis 
was published by Safouris et al. (2008) concluding that the interruption of the jet activity was 
brief, no more than a small percent of the actual age of the whole 
source. Nevertheless, based on new low-frequency observations with the GMRT array, Machalski et 
al. (2010) has repeated the dynamical age analysis but applied to the 
opposite lobes in both the outer and the inner pairs.

The source location on the southern sky hemisphere means that it is not covered by the large 
radio surveys on the northern sky: VLSS, NVSS, FIRST, WENSS. 
The observational data at 150 {\rm MHz} and 843 {\rm MHz} from similar southern surveys - TGSS 
(in progress *) and SUMSS (Bock et al. 1999), respectively - are not 
easily available for the source with LAS over 8.7 {\rm arc min} on the sky.

In the same way as for J1453$+$3308, I calculated the spectra predicted with three 
models, KDA, KDA\,EXT, and the Machalski et al. (2010) model (hereafter MJK2010), 
and compared them to the flux densities measured at six frequencies from 160 {\rm MHz} up 
to 4860 {\rm MHz} adopted 
from their Table~2. The MJK2010 model for the outer lobes assumes a significant 
fraction of thermal particles in the lobes ($k'$=10), thus also non-relativistic equations of state, i.e. $\Gamma_{c}$=$\Gamma_{B}$=$\Gamma_{a}$=5/3. The arbitrary mean 
values of the model parameters are taken as $\alpha_{inj}$=0.541, $t$=132 {\rm Myr}, 
$Q_{jet}$=1.12$\times$$10^{38}$ {\rm W}, and $\rho_{0}$=5.55$\times$$10^{-23}$ kg/m$^{3}$. 
The spectra resulting from the predicted flux densities $S_{\rm MOD}\,(KDA)$ and
$S_{\rm MOD}\,(KDA\,EXT)$ and the goodness of the fit for each model are given in Table~7 and 
are shown in Figure\,18. Also for this source, the KDA\,EXT model reproduces the observed spectrum 
better then the continuous injection models.

It is easily noticeable that in both KDA models (Table~7) the resultant $\chi^{2}_{red}$ are 
nearly equal, although the models have significantly different values of their free 
parameters. This illustrates the problem with the determination of the best KDA solution for a model 
having a large set of input parameters. The spectra predicted for the fiducial source 
analysed in Section~3.3 clearly show that the shape in a limited frequency range can be very 
similar (or almost identical). In spite of completely different values of the model 
parameters (cf. Figure~11 and 12).

\begin{table}[here]
\caption{\small Flux densities resulting from the KDA and KDA\,EXT models for J1548$-$3216 and their goodness of fit to the observed data.}
\footnotesize
\begin{center}
\begin{tabular}{@{}rrrrr}
\hline
$\nu_{0}$ & $S_{\nu_{0}}\pm\Delta S_{\nu_{0}}$ & $S_{MOD}$ & $S_{MOD}$ & $S_{MOD}$ \\
         & & {\rm MJK2010} & KDA & KDA\,EXT \\
\hline
\hline
160 & \,\,\,\,\,\,\,8400\,$\pm$\,840 & 8174 & 8465 & 8391 \\
334 & \,\,\,\,\,\,\,4737\,$\pm$\,710 & 4909 & 4838 & 4975 \\  
619 & \,\,\,\,\,\,\,3141\,$\pm$\,252 & 3064 & 2983 & 3096 \\   
1384 & \,\,\,\,1733\,$\pm$\,87 & 1561 & 1544 & 1577 \\  
2495 & \,\,\,\,\,\,\,963\,$\pm$\,30 & 920 & 928 & 917 \\   
4860 & \,\,\,\,\,\,\,415\,$\pm$\,42 & 492 & 504 & 476 \\   
\hline
$\chi^{2}_{red}$ &  & 3.18 & 3.19 & 2.61 \\ 
\hline
\end{tabular}
\end{center}

\end{table}

(*) http://tgssadr.strw.leidenuniv.nl/doku.php

\begin{figure}[here]
\includegraphics[width=10cm, height=10cm]{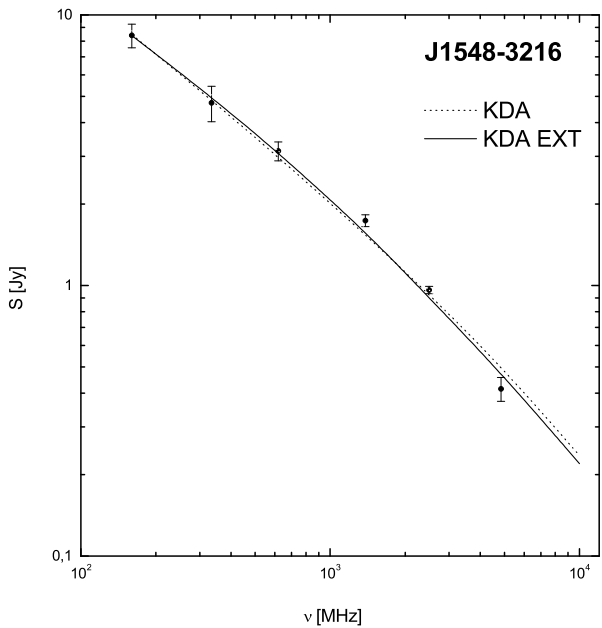}
\caption{As in Fig.\,13, but for J1548$-$3216.}
\label{18}
\end{figure}

\begin{table*}[t]
\caption{Physical parameters of the sources from the sample derived from the best KDA fits for the entire available radio spectrum (upper lines) and KDA\,EXT fits (bottom lines).}
\label{table:2}
\begin{tabular}{lllllll}
\hline \hline
Name & 3C184 & 3C217 & 3C438 &  J1428$+$3938 & J1453$+$3308 & J1548$-$3216  \\
\hline
t$_{\rm KDA}$ [Myr] & 0.78 & 1.87 & 8.5 & 66 & 75 & 55 \\
t$_{\rm KDA EXT}$ [Myr] & 0.62 & 1.70 & 13.8 & 158 & 91 & 79 \\
t$_{\rm br}$ [Myr] & 0.49 & 1.10 & 12.9 & 145 & 82 & 67 \\
\hline
$\alpha_{\rm inj}$ & 0.58 & 0.59 & 0.61 & 0.84 & 0.59 & 0.64 \\
& 0.55 & 0.53 & 0.54 & 0.58 & 0.51 & 0.60 \\
$Q_{\rm jet}$ [W] & 1.3$\times$10$^{39}$ & 2.0$\times$10$^{39}$ & 2.9$\times$10$^{38}$ & 4.1$\times$10$^{39}$ & 9.8$\times$10$^{37}$ & 8.1$\times$10$^{37}$ \\
& 1.6$\times$10$^{39}$ & 1.4$\times$10$^{39}$ & 2.0$\times$10$^{38}$ & 7.1$\times$10$^{37}$  & 5.9$\times$10$^{37}$ & 4.0$\times$10$^{37}$ \\
$\rho_{0}$ [kg/m$^{3}$] & 1.2$\times$10$^{-22}$ & 3.3$\times$10$^{-23}$ & 9.7$\times$10$^{-23}$ & 2.2$\times$10$^{-23}$ & 7.3$\times$10$^{-24}$ & 6.8$\times$10$^{-24}$ \\
& 6.3$\times$10$^{-23}$ & 1.3$\times$10$^{-23}$ & 2.6$\times$10$^{-22}$ & 0.68$\times$10$^{-22}$ & 0.56$\times$10$^{-23}$ & 0.85$\times$10$^{-23}$ \\
$p_{\rm c}$ [N/m$^{2}$] & 1.3$\times$10$^{-9}$ & 7.0$\times$10$^{-11}$ & 2.2$\times$10$^{-11}$ & 4.1$\times$10$^{-13}$ & 4.4$\times$10$^{-14}$ & 6.5$\times$10$^{-14}$ \\
& 8.4$\times$10$^{-10}$ & 3.2$\times$10$^{-11}$ & 2.0$\times$10$^{-11}$ & 0.89$\times$10$^{-13}$ & 2.6$\times$10$^{-14}$ & 0.44$\times$10$^{-13}$ \\
$U_{\rm c}$ [J] & 3.3$\times$10$^{52}$ & 2.6$\times$10$^{52}$ & 7.8$\times$10$^{52}$ & 4.3$\times$10$^{54}$ & 2.3$\times$10$^{53}$ & 1.4$\times$10$^{53}$  \\
& 1.7$\times$10$^{52}$  & 6.2$\times$10$^{52}$ & 2.2$\times$10$^{52}$ & 3.4$\times$10$^{53}$ & 2.0$\times$10$^{53}$ & 8.7$\times$10$^{52}$ \\
$B_{\rm eq}$ [nT] & 30 & 7.1 & 4.0 & 0.58 & 0.18 & 0.22 \\
& 24 & 4.8 & 3.7 & 0.25 & 0.13 & 0.18 \\
$v_{\rm h}/c$ & 0.036 & 0.068 & 0.017 & 0.018 & 0.023 & 0.024 \\
& 0.043 & 0.034 & 0.004 & 0.008 & 0.017 & 0.013 \\
\hline
\end{tabular}
\end{table*}

\section{Discussion}
~
The author is aware that the proposed model of dynamics of the 
FR-II$-$type radio sources with retained central activity 
is based on the widely used original KDA model, which assumes the steady and self-similar 
flow of energy caused by a supersonic jet. 
This assumption is reasonable in a certain ambient medium density distribution (Falle 
1991). If the jet ceases, the self-similarity of the flow appears aimless.  
Nevertheless, it should be noted that there are numerous sources for which the 
spectral aging is not yet very strong, i.e. not as evident as in the case of the typical 
radio relics (Murgia et al. 2011). The solutions and plots presented in Section\,4 
indeed suggest that the KDA\,EXT model is still useful for those sources whose jets have 
stopped relatively recently, i.e. the ($t-t_{br}$)/$t_{br}$ ratios are low (Table~8). 

As expected, in the case of two sample DDRGs (J1453+3308 and J1548$-$3216) the impact of the
renewed radio lobes interacting with the old ones is evident and may be responsible
for relatively limited spectral aging of the source. This translates into the obtained results:
the KDA\,EXT model is not statistically superior in these cases, and its goodness of fit to the 
observed data is similar to that resulting from the KDA model.

Moreover, I made some attempts to fit the extreme steepened KDA models 
(with assumed $\alpha_{\rm inj}$ values in the range of 0.8 - 1.2) to the
observed data for J1428$+$3309 and two DDRG sources. As a result, most of these fits, 
although formally best in terms of $\chi^{2}_{red}$ and rather well reproducing observations,
provide very young source ages. However, in some of these solutions 
evidently non-physical source parameters were found, i.e. the speed of the lobe expansion 
significantly greater than the speed of light, c.  
For this reason the obtained results are not included in the plots, and
the less steepened solutions (with much lower goodness of fit to the observations) are
presented instead. It can then be concluded that for these sources it is not possible to fit
a reliable KDA model with very steep spectrum (as could be expected by only taking into account the
high-frequency parts of the spectra) because their spectra are both relatively flat at the low frequencies 
and very steepened at the higher ones. This provides the additional argument for the need to apply the KDA\,EXT
model (as the one that reproduces the real data much better).

It is easy to notice that the KDA\,EXT models do not satisfactorily reproduce the bent spectra of
these sources; there are some observational data points not lying within the model
lines, or even favouring the plots of KDA model (cf. Figures\,17 and 18). Other possible
reasons for the inconsistency of this model/observations may be a variation
in the rate of particles transported in the jets (not implemented in the KDA\,EXT), or the
situation when the jet's activity is not simply stopped, but undergoes a temporary weakening
(and then resumes). In addition, the KDA\,EXT model does not include the possible unknown initial energy
distribution in the lobe's head, different from that assumed in the KDA model and demanding a new
approach to the understanding of the physics of these sources. However, it needs to be
emphasized that the range of available observations for J1428$+$3938, J1453+3308, and
J1548$-$3216 is relatively narrow when compared to the 3C sources, resulting in
not very reliable fits of both models.

It is necessary to take into account the inaccuracy of determining the initial KDA model's parameters 
obtained for the sample sources and their further use in fitting the KDA\,EXT model. This may be due 
to the inaccurate estimation of the initial model parameters for the sources (i.e. the density of 
the surrounding media, initial particle distributions, or their geometrical parameters). 
On the other hand, the goodness of the fit to the data obtained for the particular 
models may be underestimated or overestimated owing to incorrect values of the source's radio flux 
densities and their errors, which is particularly possible when using observational data from 
old radio surveys. The application of the KDA\,EXT model with the prior knowledge of some 
source parameters (which in this case would be not regarded as its free parameter, but obtained from 
independent observations), may also improve the accuracy of the solutions. A good example is a possible 
constraint to the ambient medium density from the modern X-ray observations of the field.

Finally, the n-dimensional space of free parameters 
of the KDA/KDA\,EXT models is very large and provides multiple solutions (Brocksopp et al. 2011). 
It appears that the same model outputs may be obtained using very different initial 
(input) parameters, and very often it is impossible to specify what change in these parameters 
influences the final solution the most (Section 4.3.). It may also indicate that the classical 
approach used to perform the modelling is not accurate enough, and the other numerical methods 
(general Monte Carlo) should be applied to analyse all the parameters and find the best model 
fit to the observations. 

To sum up, the extended model for the dynamical evolution of the lobes of FR\,II$-$type radio galaxies with
terminated jet activity, KDA\,EXT, was briefly presented along with the best fits 
of this model to the observed radio spectra for the small sample of radio sources. 
The results were compared to the available KDA solutions for these sources.
The preliminary results indicate that KDA\,EXT model can provide a satisfactory solution
in the case of radio sources with presumed large ages or interrupted activity.
However, the goodness of its results in relation to the observed radio spectra is
supposed to be high only for a certain type of radio sources.

Taking into account all the above, it can be assumed that the KDA\,EXT model is an appropriate 
and promising tool for solving the problem of modelling the very steep spectra of FR\,II 
radio galaxies, though the accuracy of its solutions is not high and requires further improvements. 
The future work should focus on the proposed improvements of the KDA\,EXT model, and on searching for new 
observational data to perform the further tests of the model.
\noindent
\newline

{\bf ACKNOWLEDGEMENTS}
\newline
\newline
\normalsize 

\hspace*{0.004cm} The author express her deeper thanks to the anonymous referee for the useful 
(and very essential) comments 
on the manuscript. All the recommendations and remarks significantly contributed 
to the improvement and further development of this work. I also greatly appreciate the Language 
Editor's valuable suggestions. 

This research was supported by the State Committee of Scientific Research 
through Polish National Science Centre grant No. 2013/09/B/ST9/00599. I sincerely thank 
professor Jerzy Machalski for carefully reading the manuscript and providing many valuable comments 
on the article. 
\newline
\newline
\noindent
{\bf REFERENCES}
\newline
\newline
Barai, P., \& Wiita, P.J. 2006, {\it MNRAS}, {\bf 372}, 381\\
Belsole, E., Worrall, D.M., Hardcastle, M.J, Birkinshaw, M. 2004, {\it MNRAS}, {\bf 352}, 924\\
Begelman, M., \& Cioffi, D. 1989, {\it ApJ} {\bf 345L}, 21\\
Blandford, R. D., \& Rees M. J., 1974, {\it MNRAS} {\bf 169}, 395\\
Blundell, K.M., Rawlings, S., \& Willot, C.J. 1999, {\it AJ}, {\bf 117}, 677\\
Bock, D. C. J., Large, M. I., \& Sadler, E.M. 1999, {\it AJ}, {\bf 117}, 1578\\
Brocksopp, C., Kaiser, C.R., Schoenmakers, A.P. et al. 2011, {\it MNRAS}, {\bf 410}, 484\\
Carilli, C., Perley, R. A., Dreher, J. W. et al. 1991, {\it ApJ} {\bf 383}, 554\\ 
Cohen, A.S. et al., 2007, {\it AJ}, {\bf 134}, 1245\\
Condon, J. J., Cotton, W. D., Greisen, E. W. et al. 1998, {\it AJ}, {\bf 115}, 1693\\
Daly, R.A. 1995, {\it ApJ}, {\bf 454}, 580\\
Douglas, J.N., 1996, {\it VizieR Online Data Catalog}, 8042\\
Falle, S. A. E. G. 1991, {\it MNRAS} {\bf 250}, 581\\ 
Fanaroff, B. L., \& Riley, J. M. 1974, {\it MNRAS}, {\bf 167}, 31\\
Ficarra, A., Grueff, G., Tomassetti, G, 1985, {\it A \& AS}, {\bf 59}, 255\\
Gregory, P.C., Scott, W.K., Douglas, K., Condon, J.J., 1996, {\it ApJ Suppl. Ser.}, {\bf 103}, 427\\
Griffith, M., Heflin, M., Conner, S., 1991, {\it ApJS}, {\bf 75}, 801\\
Hales, S. E. G., Baldwin, J. E., \& Warner, P. J. 1988, {\it MNRAS}, {\bf 234}, 911\\ 
Hardcastle, M.J., Alexander, P., Pooley, G.G. at el. 1998, {MNRAS}, {\bf 296}, 445\\
Jaffe, W., Perola, G. 1973, {\it A\&A} {\bf 26}, 423\\
Kaiser, C. R., 2000, {\it A\&A}, {\bf 362}, 447\\
Kaiser, C. R., \& Alexander P. 1997, {\it MNRAS}, {\bf 286}, 215\\
Kaiser, C. R., \& Cotter G. 2002, {\it MNRAS}, {\bf 336}, 649\\
Kaiser, C. R., Dennett-Thorpe, A., \& Alexander, P. 1997, {\it MNRAS}, {\bf 292}, 723\\
Kaiser, C. R., Schoenmakers, A. P., \& Rottgering, H. J. A. 2000, {\it MNRAS}, {\bf 315}, 381\\
Kataoka, J., Stawarz, \L{} 2005, {\it ApJ}, {\bf 622}, 797\\
Kellermann, K.I., Pauliny-Toth, I.I.K.,1973, {\it AJ}, {\bf 78}, 828\\
King, I. R. 1972, {\it ApJ}, {\bf 174}, 123\\
Konar, C., Saikia, D. J., Jamrozy, M. et al. 2006, {\it MNRAS}, {\bf 372}, 693\\
K\"{u}hr, H., Nauber, U., Pauliny-Toth, I.I.K. et al. 1979, A Catalogue of radio sources, Max-Planck-Institut (MPI) für Radioastronomie\\
K\"{u}hr, H., Witzel, A., Pauliny-Toth, I.I.K., Nauber, U., 1981, {\it A \& A Suppl. Ser.}, {\bf 45}, 367\\
Kuligowska E., Jamrozy, M., Kozie\l{}-Wierzbowska, D. et al. 2009, {\it AcA}, {\bf 59}, 431\\
Laing, R.A., Peacock, J.A., 1980, {\it MNRAS}, {\bf 190}, 903\\
Leahy, J.P., Perley, R.A., 1991, {\it ApJ}, {\bf 102}, 2\\
Machalski, J. 2011, {\it MNRAS} {\bf 413}, 2429\\
Machalski, J., Jamrozy, M., Zola, S. et al. 2006, {\it A \& A}, {\bf 454}, 85M\\
Machalski, J., Chy\.zy, K. T., Stawarz, \L{}. et al. 2007, {\it A \& A} {\bf}, 43M\\ 
Machalski, J., Kozie\l{}-Wierzbowska, D., Jamrozy, M. et al. 2008, {\it ApJ}, {\bf 679}, 149\\
Machalski, J., Jamrozy M., \& Saikia, D. J. 2009, {\it MNRAS}, {\bf 395}, 812\\
Machalski, J., Jamrozy M., \& Konar C. 2010, {\it A \& A}, {\bf 510}, A8\\ 
Manolakou, K., \& Kirk J. G., 2002, {\it A \& A} {\bf 391}, 127\\
Murgia, M., Parma, P., Mack, K.-H. et al. 2011, {\it A \& A}, {\bf 526}, A148\\ 
Pacholczyk, A. G., 1970, {\it "Series of Books in Astronomy and Astrophysics, San Francisco: Freeman, 1970"}\\
Rengelink, R. B., Tang, Y., de Bruyn, A. G. et al. 1997 {\it A \& A Suppl. Ser.}, {\bf 124}, 259\\
Riley, J.M.W., Waldram, E.M., Riley, J.M., 1999, {\it MNRAS}, {\bf 306}, 31\\
Roger, R. S.; Costain, C. H.; Stewart, D. I., 1986, {\it A \& AS}, {\bf 65}, 485\\
Safouris, V., Subrahmanyan, R., Bicknell, G. et al. 2008, {\it MNRAS}, {\bf 385}, 2117\\
Saikia, D. J., Jamrozy, M., Konar, C. et al. 2010, {\it Proceedings of the 25th Texas Symposium on Relativistic Astrophysics. December 6-10, 2010. Heidelberg} {\bf 14}\\
Saripalli, L., Subrahmanyan, R., \& Udaya Shankar, N. 2003, {\it ApJ}, {\bf 590}, 181\\ 
Scheuer, P. A. G. 1974, {\it MNRAS} {\bf 166}, 513\\ 
Shklovskii, I. S., 1963, {\it SvA}, {\bf 6}, 465\\ 
Schoenmakers, A. P., de Bruyn, A. G., Rottgering, H. J. A. et al. 2000, {\it MNRAS}, {\bf 315}, 371\\ 
Vigotti M., Gregorini, L., Klein, U. et al. 1999, A \& A Suppl. Ser., 139, 359\\
Waldram, E. M., Yates, J. A., Riley, J. M. et al. 1996, {\it MNRAS}, {\bf 282}, 779\\
White, R., \& Becker R. H., 1992, {\it ApJ} {\bf 79}, 331\\
Wright, E. 2006, {\it PASP}, {\bf 118}, 1711\\
Zhang, X., Zheng, Y., Chen, H. et al. 1997, {\it A \& A Suppl. Ser.}, 121, 59\\
%\end{References}
\end{document}